\documentclass[manuscript,acmsmall,screen]{acmart}
\pdfoutput=1

\usepackage{balance}
\usepackage{wasysym}
\usepackage{xcolor}         

\usepackage{enumitem}
\usepackage{mathptmx}
\usepackage{graphicx}
\usepackage{array}
\usepackage{multirow}
\usepackage{pifont}
\usepackage{color}
\usepackage{siunitx}
\usepackage{booktabs}
\usepackage{rotating}
\usepackage{xparse}
\usepackage{blindtext}
\usepackage{listings}
\usepackage{tabularx}
\usepackage{url}
\usepackage{pifont}
\usepackage[font=small]{caption}
\usepackage{subcaption}
\usepackage{tikz}
\usepackage{makecell}

\renewcommand{\paragraph}[1]{\vspace{5pt}\noindent\textbf{#1\quad}}

\usepackage{soul}
\usepackage{xspace}

\newcommand{\etal}{et al.\@\xspace}
\newcommand{\eg}{e.g.,\@\xspace}
\newcommand{\ie}{i.e.,\@\xspace}

\newcolumntype{R}{>{\raggedleft\arraybackslash}X}
\newcolumntype{L}{>{\raggedright\arraybackslash}X}

\begin{document}
\title{Understanding Longitudinal Behaviors of Toxic Accounts on Reddit}

\author{Deepak Kumar}
\affiliation{%
  \institution{Stanford University}
  \country{USA}
}

\author{Jeff Hancock}
\affiliation{%
  \institution{Stanford University}
  \country{USA}
}

\author{Kurt Thomas}
\affiliation{%
  \institution{Google}
  \country{USA}
}

\author{Zakir Durumeric}
\affiliation{%
  \institution{Stanford University}
  \country{USA}
}

\begin{abstract}
Toxic comments are the top form of hate and harassment experienced online. While many studies have investigated the types of toxic comments posted online, the effects that such content has on people, and the impact of potential defenses, no study has captured the long-term behaviors of the accounts that post toxic comments or how toxic comments are operationalized. In this paper, we present a longitudinal measurement study of 929K~accounts that post toxic comments on Reddit over an 18~month period. Combined, these accounts posted over 14~million toxic comments that encompass insults, identity attacks, threats of violence, and sexual harassment. We explore the impact that these accounts have on Reddit, the targeting strategies that abusive accounts adopt, and the distinct patterns that distinguish classes of abusive accounts. Our analysis forms the foundation for new time-based and graph-based features that can improve automated detection of toxic behavior online and informs the nuanced interventions needed to address each class of abusive account. 
\end{abstract}

\maketitle
\section{Introduction}
\noindent\fbox{%
    \parbox{0.96\columnwidth}{%
        \textbf{Content Warning}: This paper studies toxic content online. When
        necessary for clarity, this paper directly quotes user content
        that contains offensive/hateful speech, profanity, and potentially
        triggering content related to sexual assault.
    }%
}
\\\\
Toxic comments---like insults, sexual harassment, and threats of violence---are the top form of hate and harassment experienced online~\cite{thomas2021hate}. Such toxic behavior reduces the emotional safety of targets and audiences who view the content. This can lead users to self-censor to avoid further attacks, leave online platforms altogether, and in some tragic cases, inflict self-harm~\cite{pew-2017,john2018self}. Transparency reports from Meta estimate that 0.14--0.15\% of all views on Facebook in 2021 were of toxic posts~\cite{facebook-transparency-toxic}, while Twitter reports that it removed
roughly two million accounts in the second half of 2020 due to hate and
harassment~\cite{twitter-transparency-toxic}.

Prior research into toxic comments has focused on a variety of themes including the experiences of targets~\cite{silva2016analyzing, vitak2017identifying, singh2017they}, the
characterizations of specific, large-scale events like
\#GamerGate~\cite{chatzakou2017measuring}; early warnings for how toxic
conversations escalate~\cite{conversationsawry, zhang2020quantifying, xia2020exploring}, the off-platform coordination tactics for raiding and calls to incite attacks against targets~\cite{mariconti2019you, aliapoulios2021large}, and the impact of intervention techniques such as banning accounts or entire communities~\cite{ribeiro2020does, chandrasekharan2017you, chang2019trajectories}. While these studies all paint a rich tapestry of toxic behaviors that occur online, none capture the long-term activities of \textit{abusive accounts} (i.e., accounts that post toxic comments), such as their toxicity behaviors or their impact on the platform itself. Such analysis is crucial to understanding what interventions---such as nudges, warnings, and bans---might best reduce toxicity, or how automated detection can evolve to encapsulate longitudinal account behavior.

In this work, we present the results of a longitudinal, quantitative study of
abusive accounts on Reddit that post toxic comments. Over an 18~month period, we
identified 929K~abusive accounts that posted 14~million toxic comments, and
use this perspective to study three main research questions:




\paragraph{RQ1: What is the aggregate scale and nature of toxic comments and abusive accounts on Reddit?} Abusive accounts that post at least one toxic comment make up 3.1\% of all accounts that posted to Reddit during our analysis window, with their toxic comments comprising 0.8\% of all content on Reddit. Based on a manual review of abusive accounts and their activities, we estimate that 63.4\% of toxic comments are insults, 14.2\% are identity-related attacks, and 5.5\% are threats of violence, among other classes of toxic behaviors. 
Toxic comments are highly visible on Reddit: 55.2\% of Reddit accounts post directly on a thread with a toxic comment. Unlike automated, fake accounts that solely post spam~\cite{grier2010spam}, abusive accounts readily engage in non-toxic conversations, contributing an astounding 33.3\% of all comments to Reddit. As such, simply banning abusive accounts would have substantial additional consequences to the platform.

\paragraph{RQ2: What are the unique attack patterns (e.g., mob-like coordinated attacks on a single individual) that abusive accounts use when posting toxic comments online?} In graphing the reply relationships between attackers and their 1.6M~receivers\footnote{Similar to research in intimate partner violence, we intentionally avoid the term ``victim'' to not disempower people facing abuse.}(\ie accounts who received a toxic comment as a reply), we observe three classes of attacks. The vast majority of receivers (92.8\%) experience spurious, one-off toxic interactions. These receivers of abuse rarely have existing network relationships with their attacker, suggesting that the majority of abuse on Reddit is contextual and not necessarily premeditated. However, the remaining attacks are more pernicious: 7.2\% of receivers experience \emph{repeated abuse}, where a single abusive account continuously attacks the target, often across subreddits. Another 3.3\% of receivers experience \emph{flooding}, whereby a cluster of abusive accounts simultaneously attack the target, akin to coordinated raids. 

\paragraph{RQ3: What are the classes of abusive accounts, and how do they inform more nuanced defenses against toxic behaviors?} Finally, we cluster the longitudinal behaviors and activities of abusive accounts based on their posting volume, toxicity levels, subreddit participation, and community norm violations. In the process, we identify three distinct classes of attackers.  \emph{Occasional abusers}---accounts that post just a handful of toxic comments---make up 71\% of abusive accounts and 71\% of all toxic comments. This suggests that modest interventions, such as nudges or warnings, may be effective for more than two-thirds of the toxic behaviors on Reddit.  Conversely, \emph{moderate abusers}---accounts that post a substantial
volume of toxic comments---make up another 24\% of abusive accounts.
\emph{Serial abusers}---accounts that extensively post toxic comments---make up the remaining 4.3\% of abusive accounts. These two latter classes pose a
greater threat and require more stringent interventions, however, their volume of toxicity make them potentially easier to take action on.

Combined, our findings illustrate the need for nuanced interventions in tackling varied classes of toxic accounts and unique toxicity patterns. Our analysis can also help to characterize platform-wide upheavals, whereby major world events such as the murder of George Floyd~\cite{nyt-george-floyd} trigger a wave of increased toxicity across the entire platform (Section~\ref{section:case_study}). Furthermore, our findings contribute a variety of features, such as reply relationships between abusive accounts, account toxicity trends over time, and community norm violations, that can improve existing toxic behavior detection mechanisms. Such features would transform isolated content-based classification into a holistic consideration of an account's history. To this end, we plan to release anonymized datasets to researchers on request to reproduce our analyses, develop new detection mechanisms, and further explore how toxic behaviors are operationalized online. 
\section{Background and Related Work}
In this section, we provide the necessary background and describe prior work that we build on to conduct our analysis.

\subsection{Accounts that exhibit anti-social behaviors}
Our study primarily builds on a number of quantitative and qualitative studies of accounts that exhibit anti-social behaviors online, such as trolling, bullying, and toxicity. Early work at CSCW demonstrated that ``trolling'' is not limited to just a small handful of motivated accounts, but rather that ``anyone can become a troll'' depending on a number of contextual factors, such as time of day and users' moods~\cite{cheng2017anyone}. Newer studies have focused on the accounts that post toxicity and hate speech and the adversarial nature of toxic interactions. Maity et~al. study toxic conflicts on Twitter, demonstrating how context and a predisposition to toxic behaviors can cause accounts to become \textit{repeat offenders} in terms of toxic interactions~\cite{maity2018opinion}. Most similar to our work, Mathew et~al. studied longitudinal behaviors of hateful accounts on Gab, a popular fringe social platform for ``unregulated speech'', and highlighted how hateful behaviors can grow over time~\cite{mathew2020hate}.

Other lines of work have focused on the properties of abusive accounts. For example, Ribeiro et~al.\ studied abuser properties on Twitter with the aim of detecting hateful users based on their previous comments and their place in the social graph~\cite{ribeiro2018characterizing}. Their research primarily focuses on properties of accounts, for example, identifying follower-following ratios and account age as a signal for hateful behavior. Beyond abusive behavior in aggregate, several studies have looked at case studies of abuse. For example, Hua et~al.\, identify specific properties of abusers that adversarially interact with political candidates on Twitter~\cite{hua2020towards, hua2020characterizing}. Finally, our work builds off of the methods and techniques of research that has investigated fringe hate groups and online communities, including discourse on Gab~\cite{zannettou2018gab, mathew2020hate}, Dissenter~\cite{rye2020reading}, the Manosphere~\cite{horta2020evolution}, and 4chan's politically incorrect board~\cite{papasavva2020raiders}. Our study contributes a longitudinal analysis of abuser \textit{behaviors} across their entire account history on Reddit, which reveal distinct abuse and attack patterns.

\subsection{Toxicity on Reddit}
We build off a number of Reddit-focused studies to inform our experimental design and analysis choices. Gilbert documented how a culture of masculinity on Reddit forms a toxic technoculture on \texttt{r/AskHistorians}, leaving both moderators and users subject to abuse anywhere from name-calling to ``prolonged harassment, doxxing, death threats, and rape threats.''~\cite{gilbert2020run} Other studies have focused on how toxicity plays a role in shaping norms of subcommunities on Reddit~\cite{rajadesingan2020quick,chandrasekharan2018internet} and identified that subreddits often exhibit unique macro, meso, or micro-norms, highlighting challenges in applying a broad definition of toxicity throughout the entire platform. Such subcommunity specific-norms also lead to varied experiences with the precursors and effects of toxic discussions. Xia et~al~\cite{xia2020exploring} leverage the Perspective API, a commonly used toxicity detection tool, to study how specific \textit{antecedents} of a toxic interaction, such as an accounts' prior history posting toxic comments and the subreddit context, can play a significant correlative role in predicting new toxic behaviors. Finally, several studies have focused on shifts in toxicity both on-platform and off-platform due to cross community movement following notable bans~\cite{chandrasekharan2017you, ribeiro2020does}. Our study contributes distinct classes of abusive accounts on Reddit while taking into account both a global and subcommunity-specific set of longitudinal toxicity norms.

\subsection{Defenses and mitigation strategies}
Our work is grounded in prior work in studying defenses against online toxicity. The CSCW community has contributed several design recommendations, including nudges~\cite{katsaros2022reconsidering}, providing realtime feedback on toxicity~\cite{xia2020exploring}, foregrounding norms~\cite{rajadesingan2020quick}, and outright permanent bans~\cite{kou2021punishment}. Other research has focused on predicting potentially toxic behavior based on early warning signs in conversation~\cite{conversationsawry,zhang2020quantifying}, and how such signs from conversation flow can aid in forecasting personal attacks~\cite{changtrouble}. Similarly, our work builds on recent research from the computer security community, which has recently drawn parallels between classic cybersecurity problems (e.g., for-profit cybercrime) and anti-social online abuse~\cite{thomas2021hate}. Toxic content bears the closest resemblance to spam, in that it is often
unwanted content with harmful consequences to platform users. As such, our defensive recommendations and techniques also build on work on measuring and mitigating
spam~\cite{grier2010spam,levchenko2011click} and prior work that study graph-based spam propagation~\cite{nilizadeh2017poised}.

Our study contributes additional context to existing proposed defenses by outlining which defenses may be most effective for distinct classes of abusive accounts. Our results also motivate new graph-based and time-based features for toxicity detection, which can serve as the groundwork for future research in toxicity detection. 
\section{Methodology}
\label{section:methodology}
\begin{figure*}
    \centering
    \includegraphics[width=395pt]{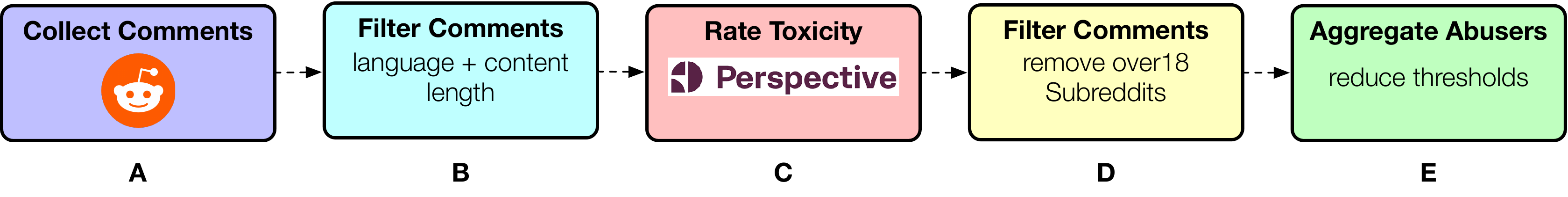}
    \caption{\textbf{Reddit Processing Pipeline}---%
        We label Reddit comments sourced from Pushshift through the Perspective
        API. We explicitly filter out comments that are not in English or
        are from subcommunities tagged as 18+. We leverage these classifications to
        identify the 14M~toxic comments and 929K~abusive accounts we study.
        }
    \label{figure:meth_figure}
\end{figure*}

In this section, we detail our methodology for capturing a longitudinal corpus of 2.2~billion Reddit comments as well as our classification techniques and thresholds for isolating abusive accounts. Figure~\ref{figure:meth_figure} shows each step in our data collection pipeline, which we use to guide our discussion.


\subsection{Collecting Reddit comments}
Our dataset consists of 18~months of comments posted on Reddit between January
2020 and June 2021. We collected a total of 2.2~billion comments via Pushshift, a
third-party API that aggregates Reddit comments and posts
(Figure~\ref{figure:meth_figure}.A)~\cite{baumgartner2020pushshift}. Each comment includes a timestamp, the username of the author, the subreddit
(\ie community) where the comment appeared, and graph data that allows us to
identify if the comment was a top-level thread (\ie the author was the original
poster), or a reply to an existing thread. From this, we re-constructed a
history of posting behavior and interactions with other accounts. At the time of writing, this constitutes the most recent data available from Pushshift.

\subsection{Filtering comments}
Prior to classification, we restricted our dataset to English comments using the
``whatlanggo'' Go
library.\footnote{\url{https://github.com/abadojack/whatlanggo}} This was
justified in part to enable manual analysis by the researchers, as well as due
to limitations with toxicity classification (described in
Section~\ref{sec:meth_model}), where existing models are trained primarily on
English text. We further omitted comments that were less than 15~characters, or
more than 300~characters in length, which is aligned with prior research on
limitations of existing toxicity models for short and very long
text~\cite{kumar2021designing}. These filters reduced our corpus to
1.8B~comments, 32.1M~accounts, and 845K~unique subreddits.


\subsection{Identifying toxic comments}
\label{sec:meth_model}

We classified the toxicity of each comment (Figure~\ref{figure:meth_figure}.C) using the Perspective
API,\footnote{\url{https://perspectiveapi.com}} a set of out-of-the-box toxicity classifiers from Google Jigsaw, which has been used extensively in prior research~\cite{hua2020characterizing,hua2020characterizing,saveski2021structure,xia2020exploring}. The Perspective API takes a comment as input and returns a score from 0--1 for several classifiers (\eg profanity, threats, identity attacks, general toxicity). As such, we had to select both a classifier(s) and a threshold to identify whether a given Reddit comment is toxic. Because the Perspective API is not explicitly trained on Reddit data, we took an additional calibration step in order to identify the best classification thresholds for our study.

To identify the best model and threshold for our context, we leveraged a public dataset that contains crowdsourced toxicity ratings for 16K~Reddit
comments~\cite{kumar2021designing}. The dataset consists of ratings from five-participants, who rated each Reddit comment on a 5-point Likert scale from ``not at all toxic'' to ``very toxic''. We consider a comment to be toxic by raters if the median score across all raters was ``moderately toxic'' or higher. We swept over each Perspective API model and threshold value (from 0--1), meaning we evaluated the precision the classifier achieved at each threshold (0.0, 0.01, 0.02, etc.) when compared to participant labels as shown in Table~\ref{table:precision_multihead}. Only the \texttt{SEVERE\_TOXICITY} classifier achieved a precision of 0.75 at a threshold of 0.9. As such, when identifying toxic comments, we filter based on if a comment has a \texttt{SEVERE\_TOXICITY} score $>$ 0.9. We note that we intentionally favor precision over recall, choosing to trade comment volume for data quality. This will underestimate the total volume of toxic comments in our corpus.

One limitation with this public dataset is that it intentionally undersamples non-toxic comments, and thus may not reflect the real precision of our threshold when applied to a truly random sample of Reddit comments. We account for this bias by manually verifying the precision of our pipeline after all classification and filtering stages, in Section~\ref{sec:meth_aggregate}. This dataset reflects a random sample of all subcommunities at Reddit at different strides of the Perspective API, and serves as a representative sample of Reddit comments to compare against.

\subsection{Removing 18+ subreddits} As part of our manual validation of the
pipeline, we observed that many comments flagged with high toxicity scores were sexually explicit and sourced from subcommunities that are tagged as 18+. These subcommunities are often sexual and consensual in nature, and do not constitute the types of attacks we are interested in studying (\eg bullying, threats, identity attacks, or sexual harassment). For example, subreddits like \texttt{r/gonewild} exist for adult participants to share consensual nude images, and posts often intentionally solicit sexual replies. In many of these cases, the classifier's high toxicity score likely does not match the intention of these community norms, and including them as toxic comments would likely negatively tag benign accounts. As such, we chose to explicitly exclude subcommunities that are tagged as 18+ from our study.  This filtering step removed 79M (3.6\%) of total comments. We stress that while harassment may occur in these subcommunities, including them and unintended false positives would negatively affect the quality of our results across the platform given our longitudinal, platform-wide approach to measurements.

\begin{table}
    \centering
    \begin{tabularx}{\columnwidth}{Xrrr}
        \toprule
        Classifier  &   Threshold   &   Precision   &   F1 \\
        \midrule
        \texttt{IDENTITY\_ATTACK}	&	0.9	&	0.62	&	0.02\\
        \texttt{INSULT}	&	0.9	&	0.53	&	0.11\\
        \texttt{TOXICITY}	&	0.9	&	0.51	&	0.24\\
        \texttt{SEVERE\_TOXICITY}	&	\textbf{0.9}	&	\textbf{0.75}	&	0.02\\
        \texttt{THREAT}	&	0.9	&	0.43	&	0.06\\
        \bottomrule
    \end{tabularx}
    \caption{\textbf{Optimal Perspective API Thresholds for Ground Truth Toxic Comments}---%
        The thresholds that maximize precision for each Perspective API
        classifier on Reddit are all 0.9 or higher, with only one
        classifier, \texttt{SEVERE\_TOXICITY}, achieving an acceptable precision
        for our measurements.
	}
    \label{table:precision_multihead}
\end{table}

\subsection{Aggregating abusive accounts}
\label{sec:meth_aggregate}
\begin{table}
    \centering
    \small
    \begin{tabularx}{\columnwidth}{Xrrr}
        \toprule
        Threshold   &   Nonabuser Prec. &  Abuser Prec. &   Comment \% \\
        \midrule
        0.5 &   0.19    &   0.56 &  64M (3.9\%) \\
        0.7 &   0.22    &   0.68 &  27M (1.7\%) \\
        \textbf{0.8}    &   0.29    &   \textbf{0.72}   & \textbf{14M (0.8\%)} \\
        0.9 &   0.0    &    0.79  &  1.7M (0.09\%) \\
        \bottomrule
    \end{tabularx}
    \caption{\textbf{Identifying Thresholds for Comments From Abusive Accounts}---%
        Reducing the toxicity threshold for comments posted by abusive accounts
        increases the volume of comments available to analyze while maintaining
        overall precision.
    }
    \label{table:abusive_threshold}
\end{table}

In the final stage of our methodology, we leveraged our corpus of labeled,
filtered comments to identify ``abusive'' accounts
(Figure~\ref{figure:meth_figure}.E). We categorized an account as abusive if it ever posted a comment with a \texttt{SEVERE\_TOXICITY} score $> 0.9$, yielding our final dataset with a total of 929K abusive accounts. We evaluated increasing the threshold (i.e., requiring abusive accounts to post $n$ comments between 1--5), but found it did not change our results (Appendix~\ref{section:abuser_thresholds}). However, by only
considering comments from these abusive accounts that meet our strict threshold of $0.9$, we significantly reduce the volume of comments available to analyze to just 1.7M (0.09\%) comments.

Given this potentially low comment volume, we next examined whether we could adopt a lower threshold for comments, conditioned on the account having posted at least one comment with high toxicity.\footnote{We demonstrate how stricter thresholds (e.g., posting at least three highly toxic comments) has little effect on the results in Appendix~\ref{section:abuser_thresholds}.} Our hypothesis was that if an account engages in toxic behavior, it is likely that some of their other comments may also be toxic. To evaluate this hypothesis, we randomly sampled 200~comments from abusive accounts and 200~comments from nonabusive accounts at each of four severe toxicity thresholds: $0.5$, $0.7$, $0.8$, and $0.9$.\footnote{For this experiment, two expert raters classified each comment as toxic or not using the definition provided by Google Jigsaw: ``a rude, disrespectful, or unreasonable comment that is likely to make you leave a discussion.'' This avoided the stratified sampling bias present in the labeled Reddit corpus.} We manually labeled each comment for toxicity and measured how well each threshold performed on our manual sample (Table~\ref{table:abusive_threshold}).

For comments from known nonabusive accounts, all thresholds performed poorly,
reaching only a maximum precision of 0.29 at a decision threshold of $0.8$
(nonabusive accounts posted no comments higher than $0.9$ by definition). For
comments posted by known abusive accounts, performance was significantly higher,
achieving a 0.72~precision at a threshold of $0.8$, and a maximum
precision of 0.79~at a threshold of $0.9$. Based on this analysis, we expand our corpus of toxic comments to include all comments---predicated on originating from an abusive account---that meet a threshold of \texttt{SEVERE\_TOXICITY} score $>0.8$. After this secondary threshold, our final dataset consists of 929K~abusive accounts and 14M~toxic comments, as well as 28.9M~nonabusive accounts and 1.6B~nonabusive comments.

\subsection{Limitations} We caution that our strategy for filtering and classification is not a perfect indicator of toxic behaviors on Reddit. Part of the motivation for this work is to identify contextual features that might improve toxicity classification, and yet, in order to do so, we need to begin with what existing classifiers are able to identify. We may omit some toxic behaviors due to false negatives, and given that our final per-comment precision is 0.72, we may also include some nonabusive comments in our final dataset. We stress that online hate and harassment is a nascent problem, made additionally challenging by differing opinions on what constitutes toxic content ~\cite{kumar2021designing,gordon2021disagreement}. Our final precision numbers are consistent and at times times, stricter with prior research in this area~\cite{hua2020towards,hua2020characterizing,saveski2021structure,rajadesingan2020quick,xia2020exploring}, and, as we find, still provide significant signal for large-scale measurement analysis.

\subsection{Ethical considerations} Without proper care, targets of abuse or the abusers themselves might be inadvertently harmed by our study. To mitigate these risks, we never interact with accounts, we never attempt to deanonymize receivers of abuse or the abusive accounts themselves, and we never report accounts to the platform due to the risk of unintended false positives. Furthermore, we note that our dataset is constructed off of an existing third-party source; we are only augmenting this existing dataset with toxicity labels using a standard approach (Perspective API) which is publicly available. We plan to release our labeled datasets to researchers by request, and we will remove any personally identifiable information (\eg account names) before release.

\section{RQ1: The Scale and Nature of Toxicity on Reddit}
\label{sec:abusers_in_aggregate}

We begin with an aggregate analysis of the 929K~abusive accounts and 14M~toxic comments they post to Reddit. We study the volume of toxic content, the subcommunities toxic content appears in, and the most prevalent types of toxic content (\eg bullying, identity attacks, threats).

\begin{figure}
    \centering
    \vspace{-10pt}
    \includegraphics[width=0.7\columnwidth]{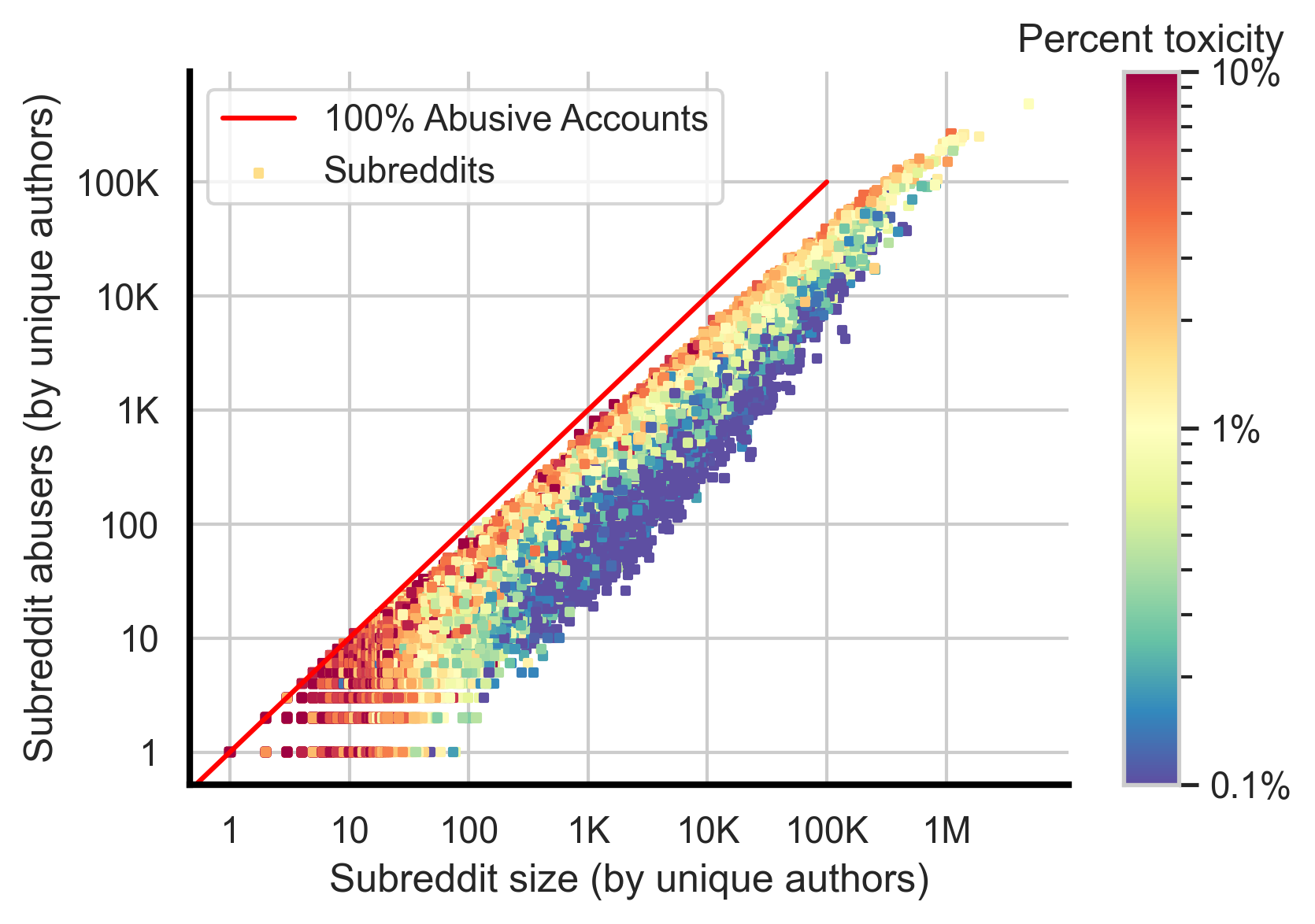}
    \caption{\textbf{Subreddit Hotspots}---%
        We show subreddits by the number of abusive accounts as well as the size of the subreddit. Some subreddits (i.e., those in the red band) serve as hotspots for toxic behaviors, hosting both high fractions of abusive accounts and large amounts of toxic content.
    }
    \label{fig:subreddit_hotspots}
\end{figure}

\subsection{Toxicity across the platform}
Accounts that post at least one toxic comment represent 3.1\% of all accounts that post comments on Reddit. Despite their relatively small presence compared to nonabusive accounts, abusive accounts play an active and outsized role on the platform. Abusive accounts post 559M~comments during our study period, which amounts to 33.3\% of all comments posted to Reddit during that time. Of these comments, 14M (2.9\%) were toxic.  Across all of Reddit, 0.8\% of all comments are toxic. We note that while abusive accounts do post a large volume of comments, simply posting significant comments does not correlate with toxic behavior ($r=0.01$, $p<0.01$), highlighting that the proclivity to post toxic comments is not simply a product of heavily using the platform. The harm caused by toxic comments is amplified when viewed by thousands of other users that engage with threads where the toxic comments appear. In total, 15M accounts (55.2\%) participate directly in a thread where a toxic comment is posted, suggesting that toxic comments are not easily avoided.

\subsection{Toxicity per individual subcommunity}
A significant number of subreddits are affected by abusive accounts and toxic
comments: 146,831 (63.4\%) subreddits have participation by abusive accounts, of
which 51K (22\%) contain at least one toxic comment. These subreddits are
typically the ones with the most activity, and 100\% of highly active subreddits
(i.e., more than 100K~comments) contain toxic comments.
Figure~\ref{fig:subreddit_hotspots} shows the fraction of abusive accounts in
subreddits, with the gradient representing the fraction of comments in those
subreddits that are toxic and the line at $y=x$ indicating the subreddit
consists solely of abusive accounts. Subreddits that fall into the red band are
\emph{hotspots} of toxic activity, which tend to be smaller subreddits that
contain tens or hundreds of unique accounts (\ie the bottom left corner of the
graph.) Such subcommunities tend to serve a niche user base (\eg
\texttt{r/FortniteBad}, which is a subcommunity where accounts share hateful
memes targeted towards the video game Fortnite), and in 1\% of cases,
consist entirely of abusive accounts.  However, even some highly active subreddits can consist of upwards of 30\% toxic content, highlighting that even outside of
hotspots of toxic activity, toxic comments are a potentially pervasive part of Reddit.

\begin{figure*}
    \centering
    \begin{subfigure}[h]{0.49\linewidth}
    \includegraphics[width=\columnwidth]{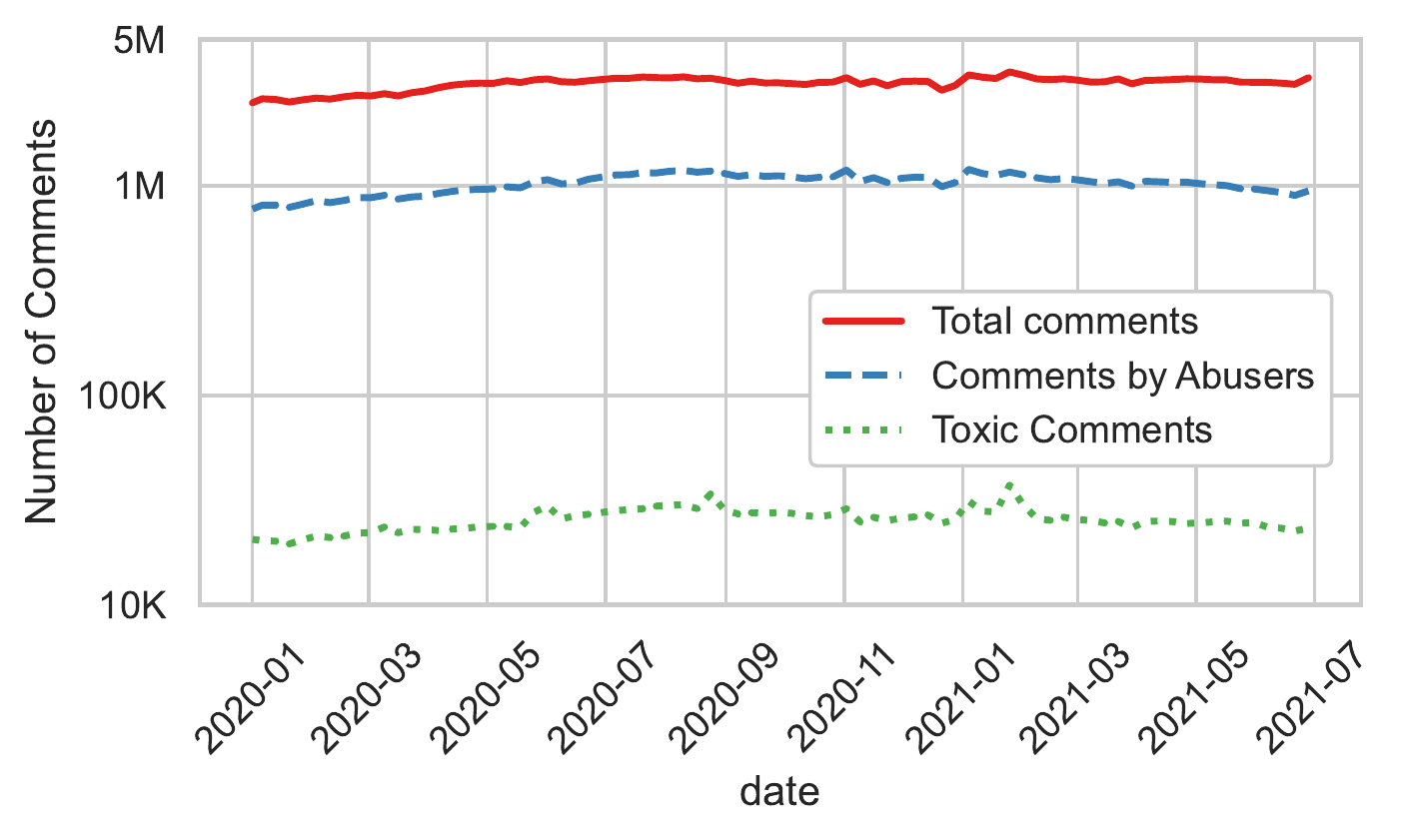}
    \vspace*{-8mm}
    \caption{\textbf{Number of Comments by Week}}
    \label{fig:toxic_timeseries}
    \end{subfigure}
    \begin{subfigure}[h]{0.49\linewidth}
    \includegraphics[width=\columnwidth]{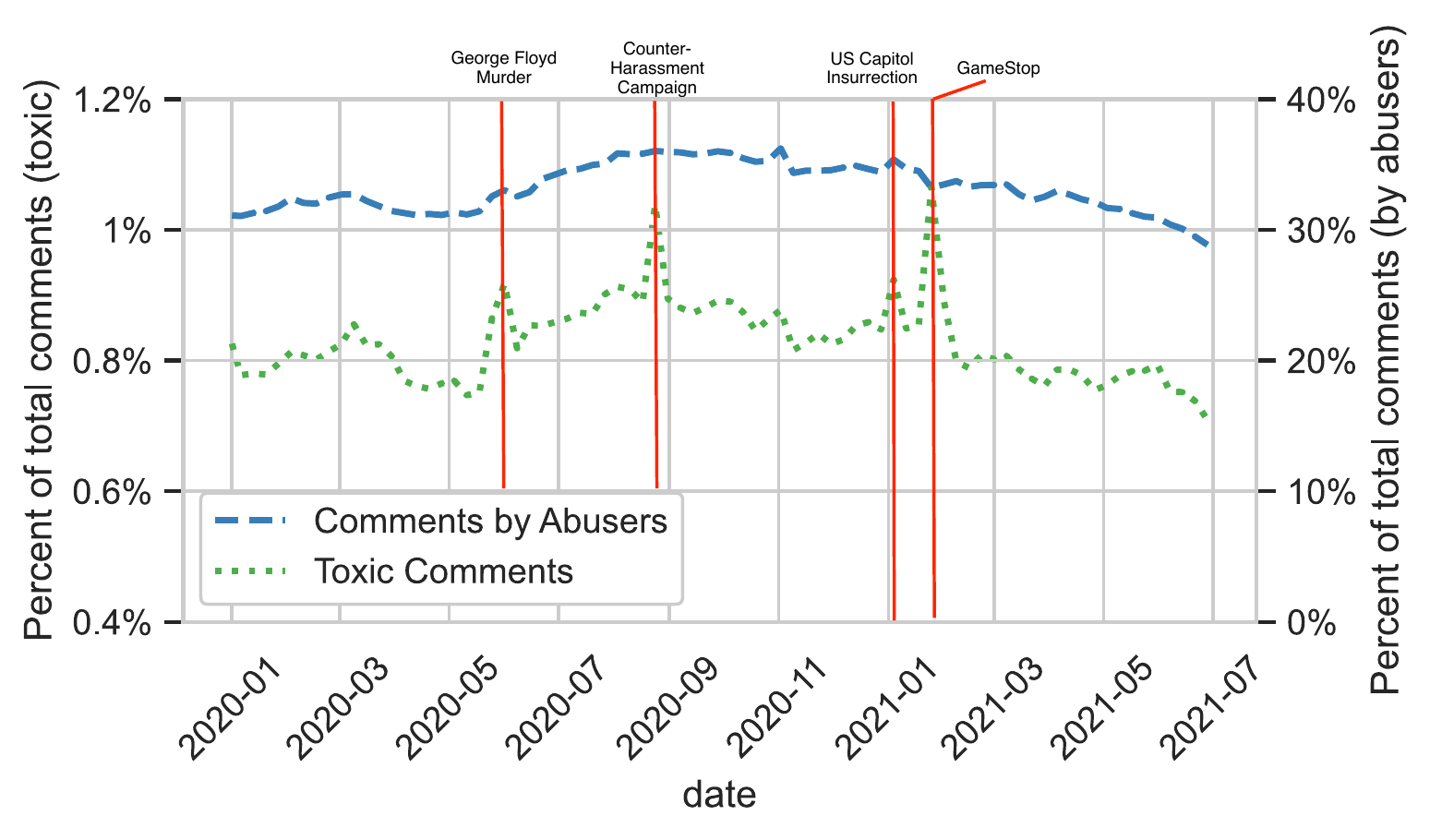}
    \caption{\textbf{Fraction of Comments by Week}}
    \label{fig:relative_volume}
    \end{subfigure}
    \caption{\textbf{Toxic Behaviors on Reddit Over Time}---%
        Toxic comments are increasing slightly over time, and account for an average of 0.8\% of all comments during our observation period. Toxic comments spike several times during our study period, typically in response to real-world events that spur significant discussion, like the murder of George Floyd in May 2020 or the January 6th US Capitol insurrection in January 2021.
    }
    \label{fig:longitudinal}
\end{figure*}

\subsection{Toxicity trends over time}
Toxic comments are regularly posted to Reddit, accounting for between
0.75--1.05\% of all comments in any given week
(Figure~\ref{fig:longitudinal}). All abusive behaviors increased over time
during our study period: in particular, the raw volume of toxic comments and
their relative presence compared to non-toxic comments has increased, which we
confirm with a Mann-Kendall trend test ($p<0.01$). The raw and relative volume
of active abusive accounts also increased over time.

We note that there are four spikes in both raw and relative volumes of toxic
behavior in our corpus. The first is between May 26 and June 5 2020, which we
manually confirmed related to posts covering the murder of George
Floyd~\cite{nyt-george-floyd}. The second spike of toxic comments occurred in
August 2020, which was due to an automated counter-campaign against a bot that
aimed to facilitate kinder language on Reddit. The third spike period occurred
on January 6th, 2021, with toxic posts largely relating to the insurrection at
the US
Capitol.\footnote{\url{https://apnews.com/article/capitol-siege-police-riots-congress-c632472d5e11063611b4a902859d49fb}}
The final spike in toxic comments came at the end of January 2021, and directly
related to the \texttt{r/WallStreetBets} takeover of the Gamestop stock on
Reddit.\footnote{\url{https://www.theverge.com/22251427/reddit-gamestop-stock-short-wallstreetbets-robinhood-wall-street}}
As a case study, we examine the first spike period in Section~\ref{section:case_study}.

\subsection{Toxic comment breakdown}
\label{section:toxic_comment_breakdown}

We manually investigated a random sample of 500~toxic comments, which we coded
into several categories of toxic behavior. We labeled each comment based on hate
and harassment categories identified in prior
work~\cite{banko2020unified,kumar2021designing}: doxxing, identity attacks,
identity misrepresentation, insults, sexual aggression, threats of violence, and
profanity. We added one additional category we find particularly prevalent on
Reddit, a ``call to leave conversation'', which typically involves the attacker
telling the target to leave the conversation, subreddit, or subcommunity. We
excluded comments from our analysis that were not relevant (\eg a false
positive or general negative sentiment). Table~\ref{table:attack_dist} shows the breakdown of attacks per category.

\begin{table}
    \centering
    \small
    \begin{tabularx}{\columnwidth}{Xrr}
        \toprule
        Category of Attack  &   \% Comments     &   Std. Error \\
        \midrule
        Insult          &   63.4\%              &   2.2\% \\
        Identity Attack &   14.2\%              &   1.6\% \\
        Call to Leave   &   12.0\%              &   1.5\% \\
        Threat          &   5.5\%               &   1\% \\
        Sexual Aggression   &   2.8\%           &   0.7\% \\
        Identity Misrepresentation  &   1.6\%   &   0.6\% \\
        Doxxing         &   0\% \\
        \midrule
        Targeted Toxicity       &   44.8\%          &   2.2\% \\
        Generalized Toxicity    &   55.2\%          &   2.2\% \\
        \bottomrule
    \end{tabularx}
    \caption{\textbf{Types of Attacks on Reddit}---%
        Attacks fall largely into two categories---attacks on authors or attacks
        on nonauthors. Attacks on authors are largely insults or calls for the
        participant to leave the community, whereas nonauthor attacks
        focus on larger identities (\eg racial, political, etc.) but are not explicitly against the author of a post or comment.
    }
    \label{table:attack_dist}
\end{table}

The majority of attacks on Reddit are insults (63.4\%), which are typically
provided in response directly to a previous commenting account or a reply to the
original poster themselves. For example, in the subreddit
\texttt{r/ShitLiberalsSay}, a community designed to mock liberal
opinions, one account wrote:

\vspace{5pt}
\begin{quote}
``I don't know what's more salty. Your mouth or your asshole. Not like there
is a big difference between them in your case, diarrhea and entitlement come
out of both ends, and they're both just as pathetic.''
\end{quote}
\vspace{5pt}

\noindent Attacks also fall into a host of other categories, including identity
attacks (14.2\%), calls to leave the conversation (12\%), and threats of
violence (5.5\%). One example of a call to leave is:

\vspace{5pt}
\begin{quote}
``...Get the fuck off this subreddit. You clearly don't really care about how
    severe NVLD can be. There's literally no fucking help out there for us.''
\end{quote}
\vspace{5pt}

\noindent We did not observe any instances of doxxing in our
analysis. However, this is likely due to our labeling and sampling mechanism, as well as our limited manual sampling (only 500 comments). More nuanced attacks like doxxing may need finer grained tools for identification~\cite{aliapoulios2021large}.

\begin{table}
    \centering
    \small
    \begin{tabularx}{\columnwidth}{Xrr}
        \toprule
        Identity Referenced &   Fraction Comments & Std. Error\\
        \midrule
        Political   &   50\%    &   9.4\% \\
        Racial      &   21\%    &   7.7\% \\
        Sexual Orientation  &   14\% &  6.6\% \\
        Religion    &   11\%    &   5.8\% \\
        Gender  &   3\%         &   3\% \\
        \bottomrule
    \end{tabularx}
    \caption{\textbf{Identity-based Attacks}---%
        Identity based attacks on Reddit span a wide variety of identities,
        including politics, race, gender, and religion. Political attacks make
        up the largest fraction of identity attacks we labeled (50\%), however,
        racial attacks and sexual orientation attacks are also highly
        prevalent.
    }
    \label{table:identity}
\end{table}

\subsection{Targets of toxic comments}
During our manual analysis, we observed two classes of the targets of toxic comments: targeted toxicity (i.e., a comment directly related to the original poster or in reply to a comment), or generalized toxicity (i.e., toxicity towards a broad group of people or a public figure). Generalized toxicity is most prevalent (55.2\%), and typically skew towards identity-based attacks (22.4\%). This is significantly larger than identity attacks launched directly on authors (6.2\%), and highlights the prevalence of undirected hate-based rhetoric on the platform. Many such attacks target race, gender, sexual orientation, political
affiliation, and a myriad of other identities. For example, on a post from
\texttt{r/TheNewRight}, a subreddit previously dedicated to extremist right-wing
content before it was banned, one account commented:

\vspace{5pt}
\begin{quote}
``They can make a better life for themselves in their own countries instead of
coming into mine and turning it into just as much of a shithole as the
shitholes they come crawling out of.''
\end{quote}
\vspace{5pt}

To better characterize the types of identities that abusive accounts
target, we labeled each identity attack comment with the identities referenced,
namely, attacks on political views, sexual orientation, racial identity,
religion identity, or gender identity (Table~\ref{table:identity}). Political
attacks on Reddit are most common, accounting for 50\% of all identity-based
attacks. This largely follows prior research where survey participants noted
that the most prominent reason they were attacked online were due to their
political beliefs~\cite{pew2021}. Racial attacks and attacks on sexual
orientation are the next most prevalent, accounting for 21\% and 14\% of
identity based attacks each. Explicit religious or gender-based attacks were the
least common on the platform, accounting for just 11\% and 3\% of attacks
respectively. Our manual sampling reveals that toxic comments on Reddit span a
variety of types of hate and harassment, tactics, and targets. Ultimately, our analysis reveals a host of varied toxic behaviors that are highly prevalent and visible across the platform.

\section{RQ2: Toxicity Patterns in the Reply Graph}
\label{section:abuser_relations}
Our analysis thus far has focused on abusive accounts and their individual
toxicity behaviors. However, these behaviors rarely happen in isolation---many
comments (56\%) are sent in reply to other comments, creating an underlying
social structure that connects accounts to one another across the platform.
Understanding these relationships can provide deeper insights into toxicity
patterns, how toxic comments are operationalized, and ultimately how toxic comments are experienced by their receivers (i.e., those that receive a toxic reply in response to their own comment). We examine the structure of these relationships by studying two social graphs based on the underlying reply-graph embedded in Reddit: one based on posting relationships between all types of accounts, and one based solely on toxic interactions between abusive accounts and the receivers of their abuse.

\subsection{Building reply graphs}
To study latent relationships between accounts, we construct reply graphs which
link participants if they interact directly with one another. Specifically, we
build a weighted, directed graph $G_n = (V_n,E_n)$ where the vertices $V_n$ are
Reddit accounts and a directed edge $e \in E_n$ represents if an account posts a
response directly to another account. Edges are weighted based on repeated
interactions between accounts. Using this graph construction as a baseline, we
build two distinct graphs which we leverage for our analysis:

\begin{enumerate}
    \item $G_1$ is generated from the entirety of the Reddit graph. Vertices
        $V_1$ represents all accounts that post a comment on the platform. This
        graph captures all reply-relationships throughout the posting period.
    \item $G_2$ is generated from solely toxic interactions. Vertices
        represent both abusive and nonabusive accounts. Edges represent a
        directed, \emph{toxic} comment-level interaction between accounts, which
        means an abusive account posted a toxic comment in response to a receiver
        account.
\end{enumerate}

\subsection{Understanding reply relationships}
\begin{figure}
    \centering
    \begin{subfigure}[h]{0.33\columnwidth}
        \includegraphics[width=\columnwidth]{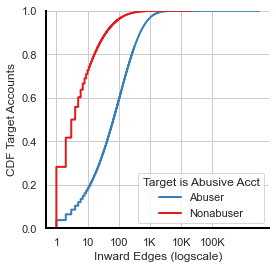}
        \caption{\textbf{Inward Edges}}
        \label{figure:in_degree}
    \end{subfigure}
    \begin{subfigure}[h]{0.33\columnwidth}
        \includegraphics[width=\columnwidth]{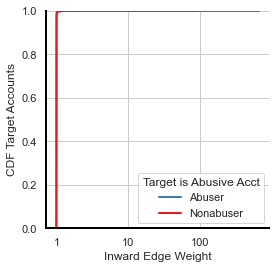}
        \caption{\textbf{Inward Edge Weight}}
        \label{figure:in_degree_weight}
    \end{subfigure}
    \caption{\textbf{Abuser and Nonabuser Connections}---%
        Abusive accounts form more connections than nonabusive accounts, largely
        due to their increased activity on the platform at large. However, both
        accounts share in identical average edge weight (median 1.0), suggesting
        that most connections formed by both types of accounts are spurious,
        one-off interactions.
    }
    \label{figure:in_degree_stats}
\end{figure}

\begin{figure}
    \centering
    \includegraphics[width=0.7\columnwidth]{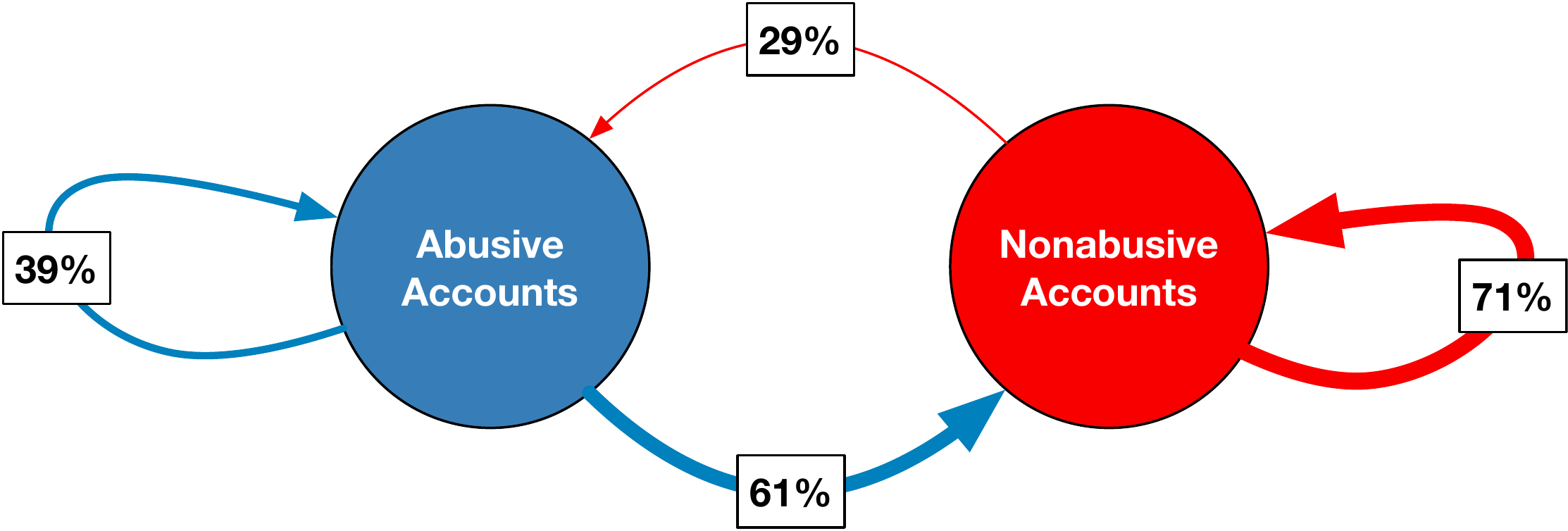}
    \caption{\textbf{Connectivity of Abusive and Nonabusive Accounts}---%
        Abusive accounts regularly interact with other abusive accounts on the
        platform, with abusive links making up for 39\% of their overall
        interactions. Conversely, nonabusive accounts interact less with
        abusive accounts, with abusive links making up an average of 29\% of
        their connections.
    }
    \label{figure:abusive_connectivity}
\end{figure}

To first understand reply-relationships formed by abusive accounts, we
investigate the structure of $G_1$, which captures the full Reddit connectivity
graph. The graph contains 22.2M~accounts and 631M~edges, of which 890K (4\%)
are abusive accounts and 21.3M (96\%) are nonabusive accounts. We note the
remaining 39K~abusive accounts in our dataset only posted toxic comments as
top-level replies to posts and never directly engaged with any other account on
the platform. We compare reply-relationships formed by abusive accounts and
nonabusive accounts to identify if abusers form fundamentally different
relationships on the platform compared to nonabusive accounts. We measure these
in three ways: the raw number of connections, the weight of those
connections, and finally, whether accounts tend to form connections to
similar accounts (\ie the graph exhibits homophily).

\paragraph{Abusive accounts interact heavily.}
Abusive accounts interact with significantly more accounts compared to
nonabusive accounts (Figure~\ref{figure:in_degree_stats}).  Abusive accounts
have a median of 67~inward connections (\eg replies from other accounts) and 63
outwards connections (\eg replies to other accounts) compared to nonabusive
accounts, who have a median 3~inward and outward connections. This aligns with
our previous observation that abusive accounts are more active on the platform
(Section~\ref{sec:abusers_in_aggregate}) and adds additional context that
abusers are also engaging in more conversations online, potentially leading to
an increased likelihood of a toxic interaction.

\paragraph{Most reply relationships are spurious.}
Although abusive and nonabusive accounts form vastly different numbers of
connections, they are similar in how deep those connections are. To measure
this, we study connection weight, which is the number of times that an account
repeatedly interacts with another account. Of the 169M~outward edges created by
abusive accounts, 147M (86.9\%) have an edge weight of 1, which are interactions
that only occur a single time during our observational period
(Figure~\ref{figure:in_degree_weight}). This is consistent with nonabusive
accounts, for which 89\% of outward edges have an edge weight of 1. The
majority of connections on Reddit, regardless of account type, are
\emph{spurious}, and are rarely repeated on the platform.

\paragraph{Homophily is low.}
All accounts at baseline exhibit small amounts of \emph{homophily}, which is a
tendency to connect with more similar accounts than dissimilar accounts. In this
context, this means that abusive accounts tend to form reply-relationships with
other abusive accounts, and nonabusive accounts tend to form reply-relationships
with other nonabusive accounts.  To capture this, we compute the assortativity
coefficient, which takes a value from -1 (indicates accounts only interact with
dissimilar accounts) to 1 (indicates accounts only interact with similar
accounts). The coefficient value is $r=0.05$, which is a small but significant
tendency for similar accounts to connect with one another, suggesting that
abusive behavior alone does not explicitly connect similar accounts together. As
such, connections to abusive accounts are similar when comparing abusive and
nonabusive accounts (Figure~\ref{figure:abusive_connectivity}). Abusive
accounts have on average 39\% of their overall connections made up by other abusive
accounts. This is marginally higher than the average fraction of connection that
abusive accounts make up for nonabusive accounts (29.6\%).

\subsection{Abusers and receivers}
\begin{figure}
    \centering
    \includegraphics[height=180pt,keepaspectratio]{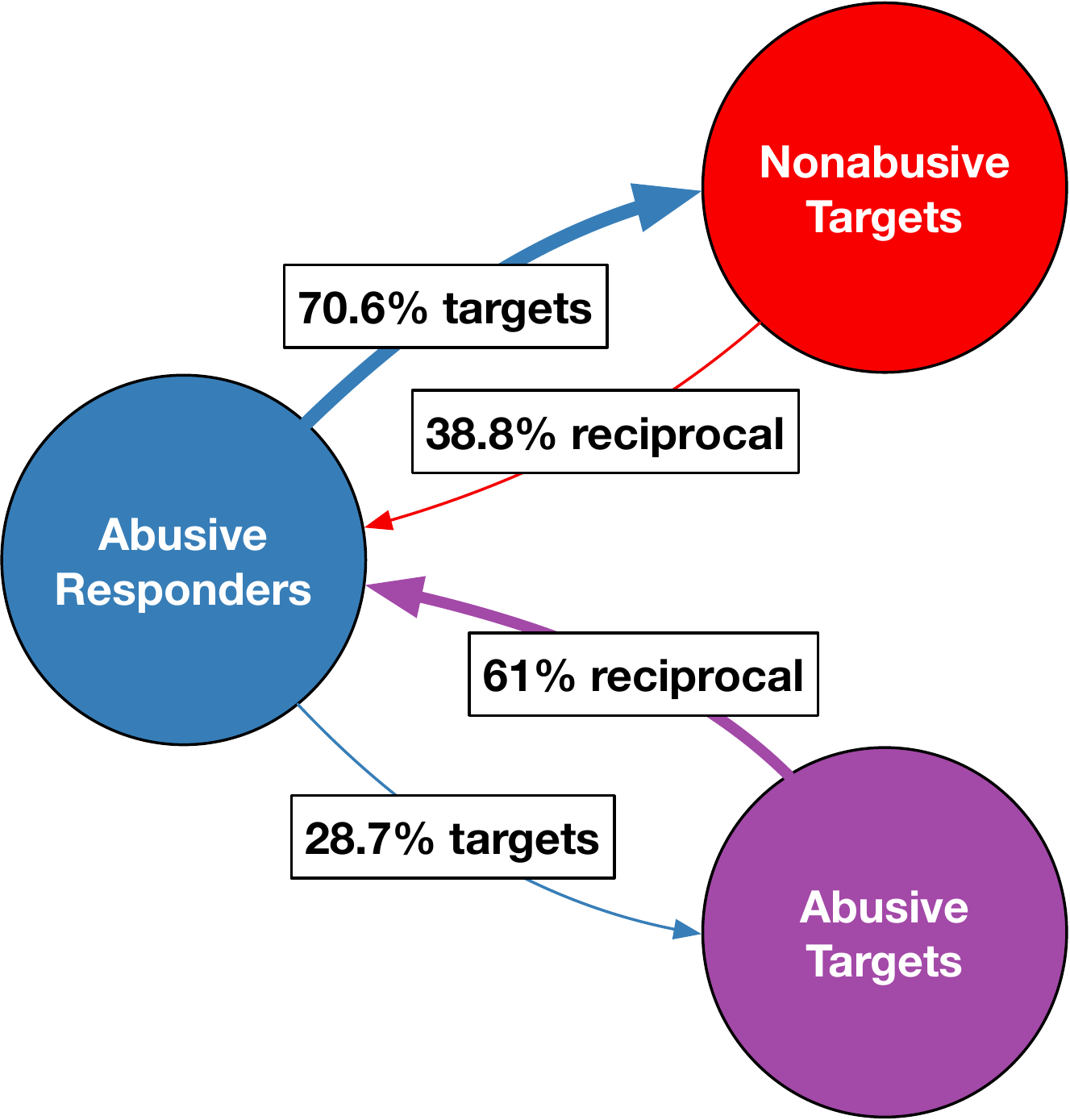}
    \caption{\textbf{Toxic Interactions}---%
        Receivers of abuse can be nonabusive accounts (70.6\% of accounts) or
        abusive accounts (28.7\%), highlighting a dual role that abusive accounts
        play in toxic interactions. Abusive accounts are more willing to extend
        toxic interactions on the platform than nonabusive accounts, with 61\%
        of abusive accounts responding to a toxic comment compared to 38.8\% of
        nonabusive accounts.
    }
    \label{fig:toxic_connections}
\end{figure}

We contrast our analysis of the aggregate reply-graph ($G_1$) with $G_2$, which
focuses strictly on toxic interactions stemmed from abusive accounts. In the
creation of this graph, we filtered out any interactions where accounts replied
in a toxic manner to \emph{themselves}, which accounted for 0.8\% of toxic
interactions. The resultant graph contains 1.8M~vertices and 4.1M~edges, of
which 1.6M (89.3\%) are receivers and 651K (36.3\%) are abusers.\footnote{We note
this does not sum to 100\% because some abusers are also receivers of abuse.} We
focus our attention on the roles that accounts play and the behaviors they
exhibit when engaging in toxic interactions.

\paragraph{Abusive accounts play dual roles.}
The majority of toxic comments are sent towards nonabusive accounts, who make up 71.3\% of all receivers. However, we note that abusive accounts can play the role of both an abuser and as a receiver of abuse
(Figure~\ref{fig:toxic_connections})---28.7\% of receivers are abusive accounts, and 460K (70.6\%) of abusive accounts play both roles in $G_2$. This aligns with our earlier result that many abusive accounts regularly interact with other abusive accounts on the platform. This dual role also highlights an underlying challenge with defending against toxic content, as account-level interventions may inadvertently harm abusive accounts when they are also receivers of abuse.

\paragraph{Abusive accounts reply to toxic comments.}
47.4\% of toxic edges have a \emph{reciprocal} connection, which means the
receiver of the toxic comment replied to the original comment
(Figure~\ref{fig:toxic_connections}). We count both toxic and non-toxic replies
as a reciprocal edge. Many abusive accounts engage in
discussion with other accounts, especially when the interactions are toxic. 421K
(54\%) of abusive accounts have reciprocal edges---this is notably different
than considering behaviors in $G_1$, where abusive accounts largely interacted
only a single time with other participants. When abusive accounts respond to
toxic comments, some (18.1\%) will occasionally respond with a toxic comment
(23.3\% of interactions), potentially escalating the toxicity of the
conversation with their reply. For example, when one abusive account posted:

\begin{quote}
    ``I just partially agreed with you, can you not read? I just said if Trump
    knew about his acts then he should of spoken about it. There is no evidence
    that trump is a rapist and a pedo you fucking retard, present it me?''
\end{quote}

\noindent The receiving abusive account replied:
\begin{quote}
    ``He's literally admitted it multiple times. You pedo apologists are fucking
    sick. You're blocked. I don't talk to pedophiles or rapists or people fucked
    enough to try to lie for them.''
\end{quote}

\noindent Our results suggest that only a minority of abusive
accounts respond in a toxic manner, adding nuance to the types of accounts
that may choose to instigate toxic interactions.

\paragraph{Most toxic interactions are one-offs.}
The overwhelming majority of toxic interactions on Reddit are one-offs, meaning
they occur one time between an abusive account and a receiver account and never
occur again. Of the 4.1M~abusive comments posted by an abusive account in
response to receiver accounts, 95.6\% are one-offs, which is larger than the
similar fraction measured from $G_1$ (83.2\%). Toxic interactions are thus often
fleeting and one-off occurrences on the platform.

\paragraph{Some abusive accounts have pre-existing relationships with abusive accounts.}
\begin{figure}
    \centering
    \includegraphics[width=0.7\columnwidth]{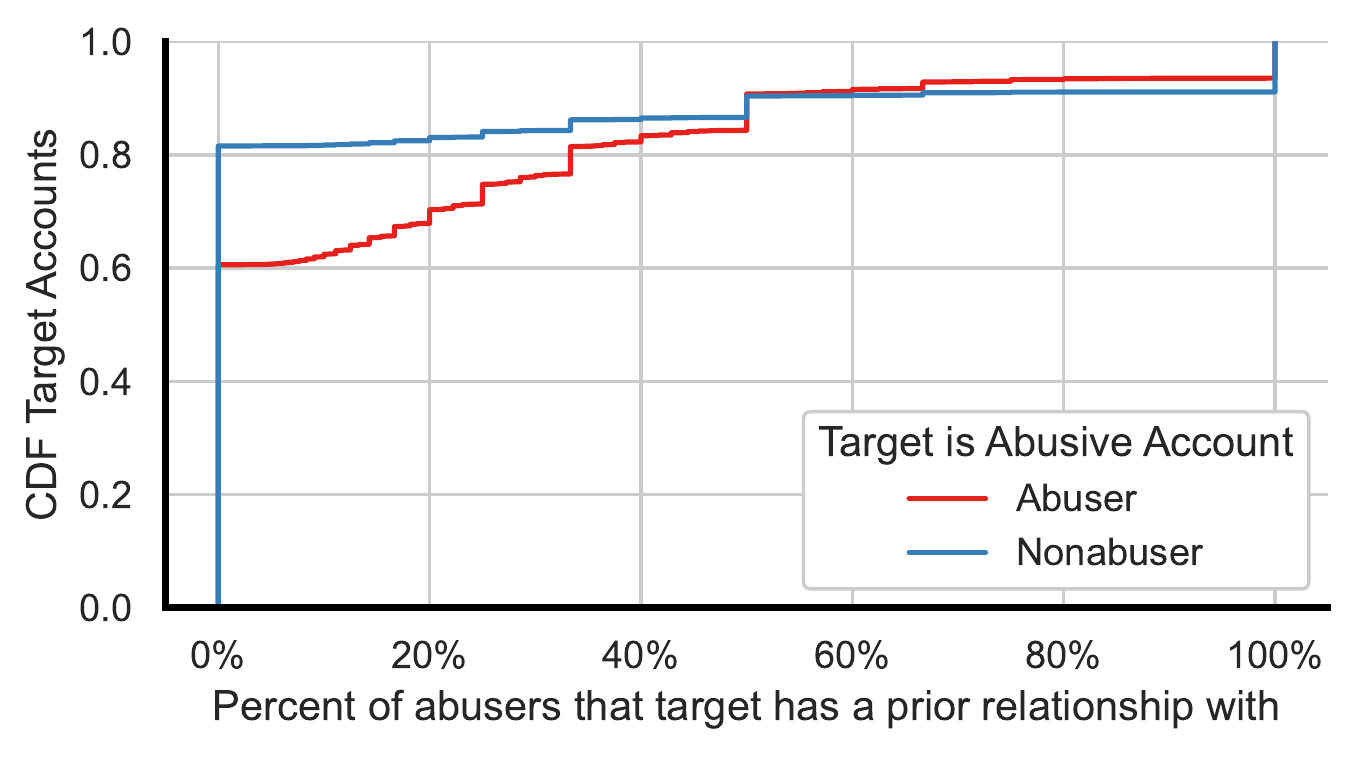}
    \caption{\textbf{Previous Relationships with Abusers}---%
        75\% of receivers do not have pre-existing relationships with their
        abusers. In contrast to this, 40\% of abusive accounts
        have pre-existing network relationships with their abusers, suggesting
        again that abuser-to-abuser communication may predicate future toxic
        interactions.
    }
    \label{fig:prev_relationships}
\end{figure}
We next identify whether receivers have existing network relationships with
abusive accounts \emph{prior} to a toxic encounter, which may inform if
underlying network features can help to predict toxic behaviors. 75\% receivers
have \emph{no} existing relationship with \emph{any} of the abusive accounts
that target them. This value is significantly lower, however, when considering
abuser receivers---40\% of abusive accounts have a pre-existing, non-toxic
relationship with their abusers. As such, toxic interactions between abusers may
be predicated by previous interaction on the platform
(Figure~\ref{fig:prev_relationships}). Excluding one-off abusive
interactions paints a different picture, which naturally increases the number of
pre-existing relationships between receivers and their abusers. When only
considering receivers with more than one abuser, 44\% have a pre-existing
relationship with at least one of their abusers, and 53\% of abuser receivers have
a pre-existing relationship with their abusers. These underlying relationships
may be useful when automatically detecting new, abusive accounts.

\begin{figure}
    \centering
    \includegraphics[width=0.7\columnwidth]{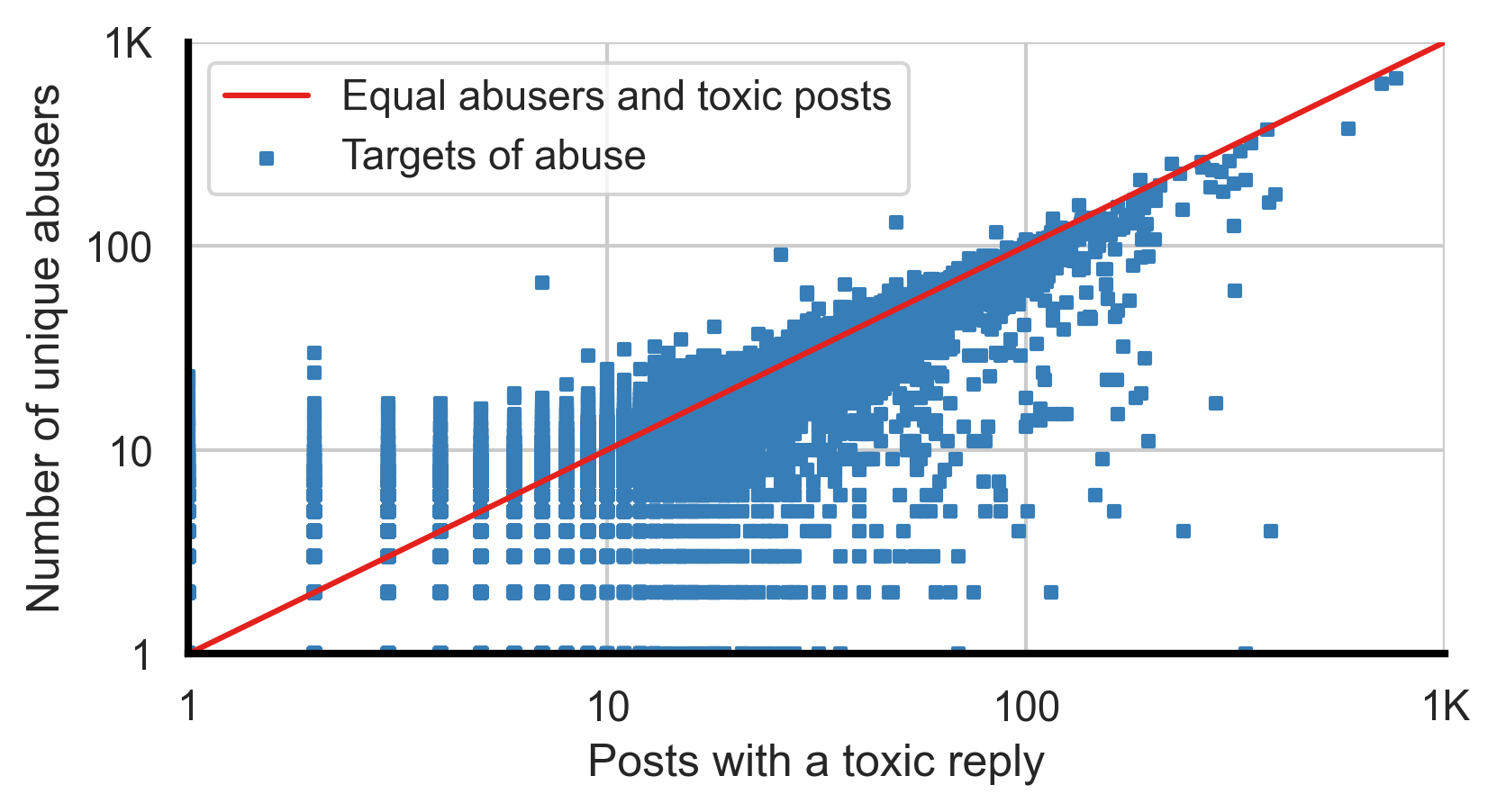}
    \caption{\textbf{Receiver Experiences}---%
        Receivers experience three distinct types of toxicity: spurious abuse
        (\eg a single abusive accounts replies to a single comment), repeated
        abuse (\eg a small handful of accounts repeatedly harass a target
        through their posting history), or flooding (\eg one comment triggers
        many toxic replies). The line at $y=x$ denotes an equal number of abuser
        and posts that trigger toxic replies. While 85.8\% of receiver accounts fall on the line, 14.2\% experience either flooding or repeated abuse.
    }
    \label{figure:target_experiences}
\end{figure}

\subsection{Receiver experiences}
Despite the fact that the majority of interactions between abusive accounts and
receiver accounts are one-offs, many receivers experience many different types of
toxicity. 59.5\% of receivers experience just one abusive interaction from
one abusive account in $G_2$, however, 40.5\% of receiver
accounts experience multiple abusive encounters during their time on the
platform. To better explain these experiences from a receiver perspective, we
measure receiver experiences by two criteria: the number of unique abusers that
send toxic comments to the receiver, and the number of posts that lead to a toxic
reply (Figure~\ref{figure:target_experiences}). The line at $y=x$ indicates an
equal number of abusers and toxic interactions. In doing this, we illuminate
three distinct toxicity patterns: spurious abuse, repeated abuse,
and flooding. We detail each below:

\paragraph{Spurious Abuse.}
92.8\% of receiver accounts experience spurious abuse, meaning each toxic
interaction that they encounter comes from a distinct abusive account. These accounts
are those that fall on the line at $y=x$ in
Figure~\ref{figure:target_experiences}.  Such experiences are spurious in nature
and have more to do with the content of the discussion rather than specific
toxicity directed towards the receiver account themselves. Protecting these receivers from unwanted toxicity is the most challenging, as they often have no prior relationships with their abusers and attacks may happen without any explicit warning.

\paragraph{Repeated Abuse.}
A significant number of receiver accounts (114K, 7.2\%) experience
\emph{repeated} abuse, which are repeated toxic comments that come from the same
abusive account. In Figure~\ref{figure:target_experiences}, these accounts fall
below the line at $y=x$. These are accounts whose comments regularly trigger a
toxic reply, but are targeted by a smaller group of abusive accounts. Most
alarmingly, abusive accounts seek out and repeatedly harass 5700 (0.5\%) of
receivers across \emph{different} subredddits.

As an example of this type of abuse, in the subreddit
\texttt{r/Syracuse\_comments}, we observe one account repeatedly harassing
another account for their support of then US President Donald Trump across
20~distinct threads. The abusive account regularly antagonizes the receiver
account for their beliefs and refers to the account in the second person,
tacitly acknowledging the abuse:

\begin{quote}
    Yes, ``news'' must never be shared. Journalists must go out and find their
    own ``news''. Pffft, Are you fucking retarded? Do you know how stupid you
    sound? Of course, it's to be expected from you.
\end{quote}

\noindent In one explicit instance, the abusive account even references their excitement
to getting to abuse the receiver account again:

\begin{quote}
    thanks for coming back for another ass kicking. It's Just so damn easy. LMAO,
    Back to the trailer. Just end it...
\end{quote}

\noindent Repeated abuse has a distinct behavioral footprint from the majority of toxic
interactions on Reddit. As such, it may be easier to defend against as it can be
more readily identified on a platform level. Despite this, we are aware of no existing
proactive protections for repeated abuse on Reddit.

\paragraph{Flooding.}
In contrast to those that experience repeated abuse, receiver accounts that fall
above the line at $y=x$ in Figure~\ref{figure:target_experiences} experience
\emph{flooding}, where a single comment may trigger many abusive interactions in
reply. In our dataset, 53K (3.3\%) receivers experience flooding, with the most
flooded receiver experiencing 74~toxic replies to a single comment. In one such
case, an account posted in the subreddit \texttt{r/JusticeServed} expressing
skepticism about Covid-19 vaccines, stating:

\begin{quote}
    Leaving this up since it's high karma, but the entire world needs to have a
    good look at vaccines. I'm not quite sure how they work but putting things
    like that in your body cannot be good for you when herbal remedies have
    worked for thousands of years. People need to start thinking for themselves
    and stop blindly trusting so called ``masters of medicine''.
\end{quote}

\noindent This comment was met with 74~distinct abusive accounts berating the author,
insulting their intelligence, and in some cases, wishing for the author's
death---as an example:

\begin{quote}
    My God you're a cunt. In a way I hope people like you stay anti-vaxx, eventually
    you'll die out.
\end{quote}

\noindent These types of flooding attacks impact a significant number of receivers and happen almost in real-time, rendering post-hoc moderation limited in its effectiveness.

\section{RQ3: Categorizing Abusive Accounts}
\label{section:behaviors}
Abusive accounts often exhibit distinct toxicity behaviors on Reddit. Our goal is to understand these behaviors and leverage them to build \emph{abuser personas}, which are  groups of abusers that share behavioral traits. Such personas can help to inform more effective defensive mechanisms or platform design choices. We focus on three distinct toxicity behaviors that we then use to build personas: abusive accounts' toxicity behaviors on the platform in aggregate, their absolute toxicity behaviors in the subcommunities they participate in, and finally, their behaviors in relation to subcommunity norms.

\subsection{Toxicity behaviors in aggregate}
\begin{figure}
    \centering
    \includegraphics[width=0.7\columnwidth]{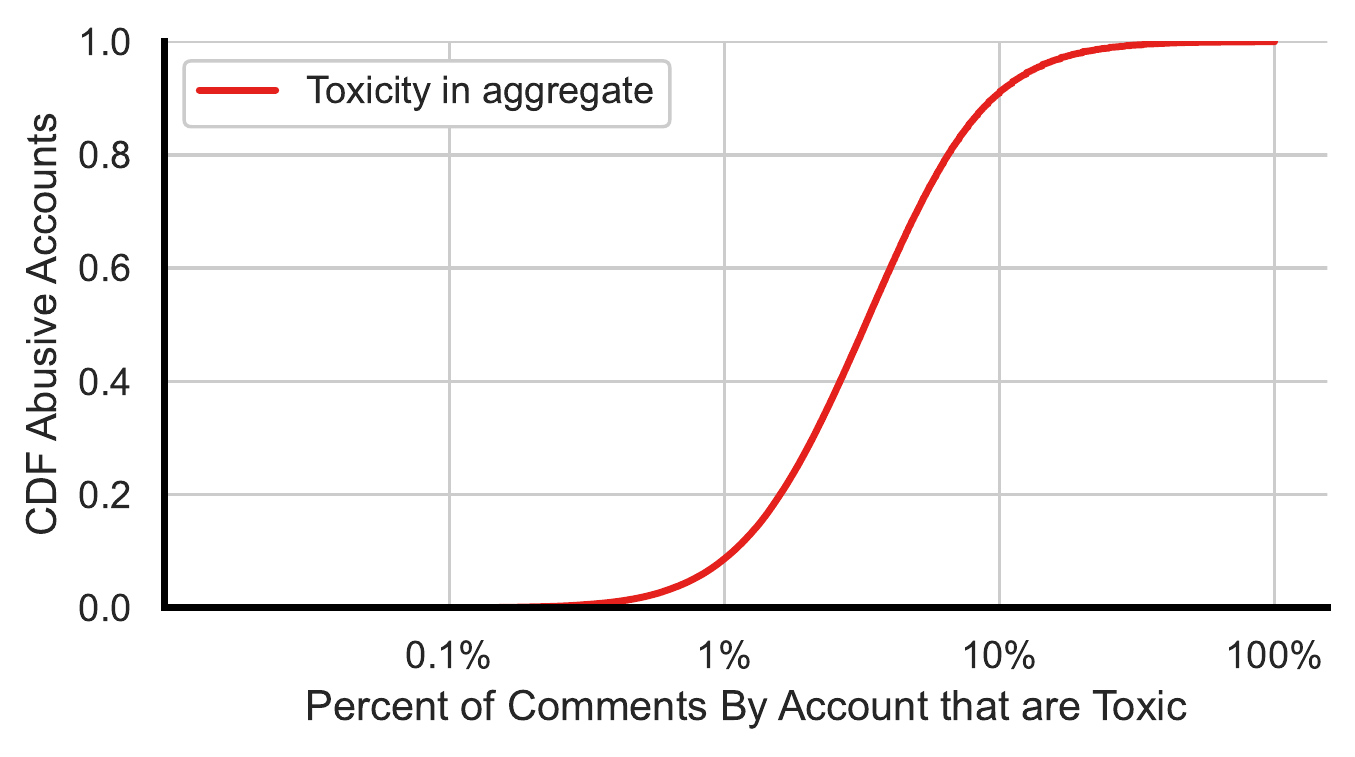}
    \caption{\textbf{Abuser Behaviors by Toxic Comments}---%
        We show a CDF of the fraction of toxic comments abusive accounts post in
        aggregate on the platform and the median amount from their
        subcommunities. Toxic comments account for a median 2.9\% of all abuser
        comments.
    }
    \label{fig:abuser_behavior_comments_cdf}
\end{figure}

\begin{figure}
    \centering
    \includegraphics[width=0.7\columnwidth]{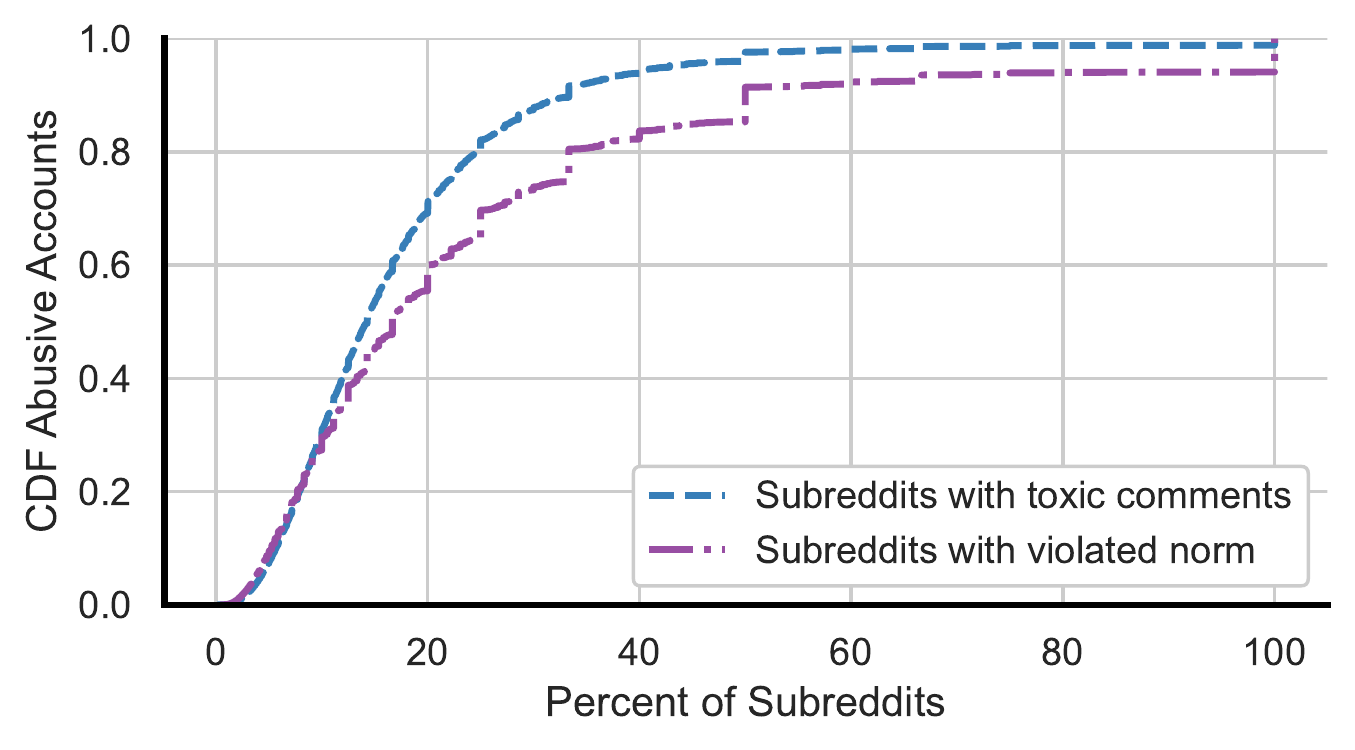}
    \caption{\textbf{Abuser Behaviors by Subcommunity}---%
        Abusers post toxic comments in a median 13\% of their subcommunities
        and violate the toxicity norms of a median 16.6\% of their social homes.
        Accounts exhibit different toxicity behaviors depending on the
        subcommunity they engage in, painting a fractured set of abusive
        behaviors that vary from account to account.
    }
    \label{fig:abuser_behavior_subreddits_cdf}
\end{figure}

\begin{figure}
    \centering
    \includegraphics[width=0.7\columnwidth]{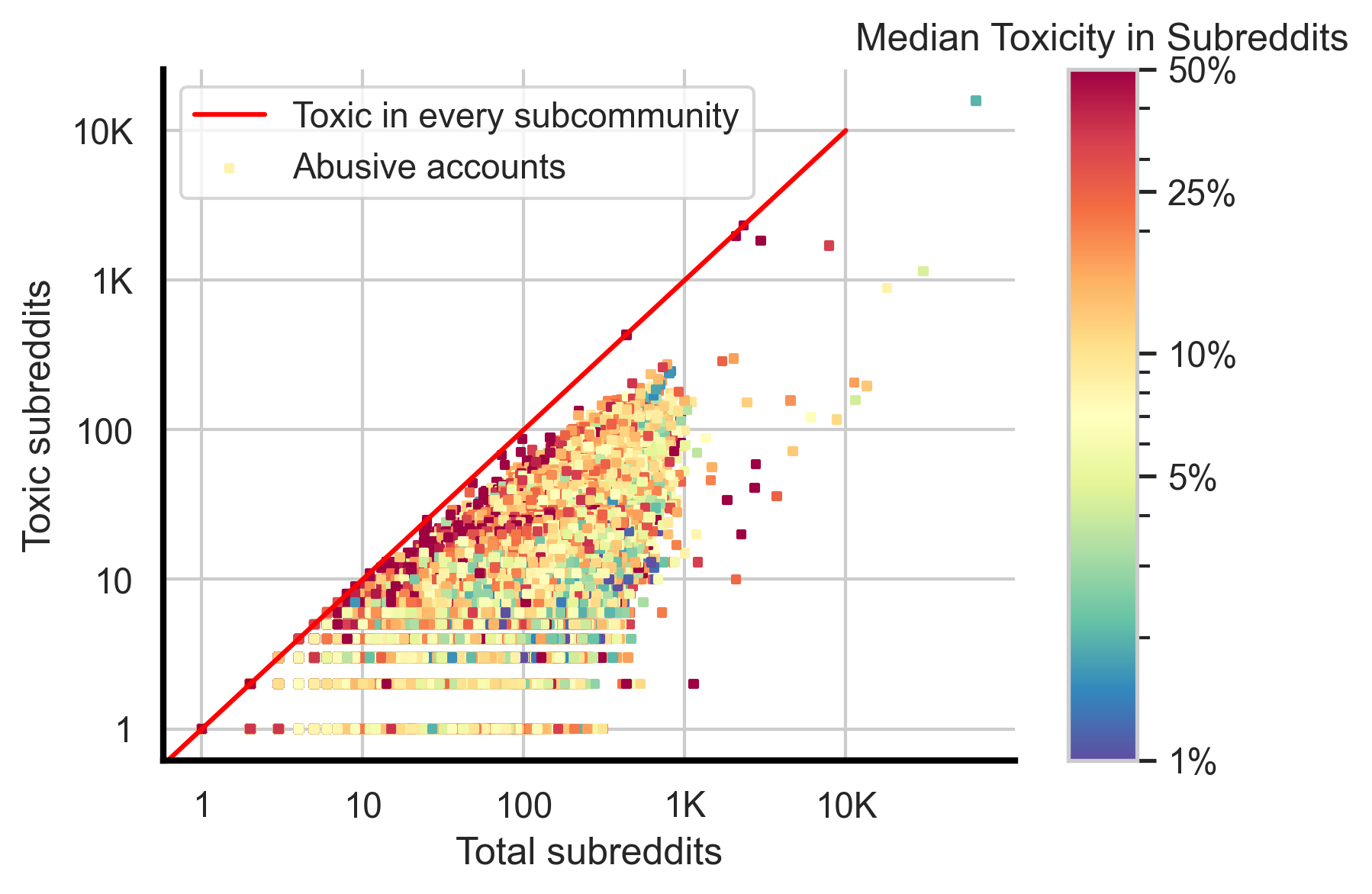}
    \caption{\textbf{Abusers in Subreddits}---%
        Most abusive accounts restrict their toxic activity to a subset of
        their communities, however, 1\% are highly toxic in nearly all the
        subcommunities they belong to. The gradient represents the median
        percent toxicity for each abusive account in their respective
        subreddits. Most abusive accounts fall into the yellow and blue portions
        of the graph, highlighting low overall toxicity in the majority of their
        communities.
    }
    \label{fig:abusers_in_subreddits}
\end{figure}

Abusive accounts post a median six~toxic comments during our study period and
toxic comments make up a median 2.9\% of all comments posted by abusive
accounts.  Figure~\ref{fig:abuser_behavior_comments_cdf} shows the distribution
of toxic behaviors per abuser. A small handful of accounts (1.3\%) exclusively
post toxic comments (100\% of their comments are toxic), many of which are
low-activity accounts, posting just a median five~comments in total.
When we consider highly active users (\ie ones that post more than
100~comments), the top 5\% of abusive accounts post only 10.7\% toxic comments.
In the most extreme case of this, 25K (4.2\%) highly active abusive accounts
post just a single toxic-message during our observation period, suggesting that
the majority of their posting history is non-toxic. On the other end, 16K
(2.6\%) of abusive accounts post at least 100~toxic comments, and toxic comments
account for at least 10\% of all comments for 18.7K (3.1\%) of accounts.

Given the rise in toxic comments posted to the platform in aggregate
(Section~\ref{sec:abusers_in_aggregate}), we also measure whether abusive
accounts increase in their individual toxic behaviors over time. We compute a
Mann-Kendall trend test for every abuser that posts at least a single comment
every week, and find that the majority of abusers (88.2\%) exhibit no change in
their toxicity behaviors over time. As such, toxicity behaviors are
\emph{stable} for most accounts and can form a foundation for building abuser
personas.

\subsection{Toxicity behaviors in subcommunities}
Even if an abusive account is highly toxic during their lifetime on the
platform, they are rarely abusive in \emph{all} of the subcommunities they post
in (Figure~\ref{fig:abuser_behavior_subreddits_cdf} and
Figure~\ref{fig:abusers_in_subreddits}). Abusive accounts comment in a median
27~subreddits overall, but only post toxic comments in a median 13\% of those
subreddits. Abusive accounts may selectively choose when or where to be toxic on
the platform, due to a myriad of factors---for example, Cheng et~al.\ found that
seeing other trolling comments had an impact on whether an account would post a
trolling comment themselves~\cite{cheng2017anyone}.

However, even when abusive accounts post toxic comments in their subreddits, it
still may account for a relatively small portion of their posting volume in each
subreddit. Indeed, when abusive accounts post a toxic comment in a
subreddit, such comments make up a median of 12.5\% of the comments they post.
As such, abusive accounts may not only be selective in
which subcommunities to post toxic comments in, but also in \emph{how toxically}
they behave in those communities. Figure~\ref{fig:abusers_in_subreddits}
represents this idea as a gradient, where the intensity of the gradient (from
blue to red) indicates an abusive account that posts significant toxic comments
in \emph{all} of their subcommunities. Most abusers fall between the blue and
yellow portions of the graph (94\%), which indicates they typically post only a small number of toxic comments in their subreddits. Yet, there are a small fraction (1\%) of
abusive accounts that choose to be highly toxic in the majority of their
subcommunities (\ie deep red points in the graph), again highlighting variance
in abusive account behaviors.

\subsection{Violating subcommunity norms}
Each subcommunity on Reddit is self-moderated, meaning each establishes its own
set of unique rules, or \emph{norms}, about what types of discussion is allowed.
As such, posting toxic comments in all communities may not explicitly violate
community norms, which a broad definition of toxicity (as we have been applying
in this paper) may not appropriately capture. In order to study abusive
behaviors in the context of community, we additionally consider how abusive
accounts violate toxicity norms defined by Rajadesingan
et~al.~\cite{rajadesingan2020quick}.

We define a subreddit's toxicity norm as the fraction of toxic content posted in
each subreddit. We restrict our analysis to subcommunities that have more than
50~comments in our dataset, and to those that exhibit a ``stable'' norm over our
measurement period, which means the toxicity norm does not change beyond 2\%
from month to month (a stability metric defined in prior
work~\cite{rajadesingan2020quick}). In addition, we filter out subcommunities
that do not have a distinctive toxicity norm, which in this context means their
toxicity norm is at least 2\% away from the average toxicity norm of all
subreddits during our study period (which was 2.4\%). In total, we identify
35,840~subreddits with stable, distinctive toxicity norms for further analysis,
of which 94.9\% of subcommunities fall below and 5.1\% fall above the platform-wide toxicity norm. We note that we only investigate a subreddit for an abusive account
if they post a comment in that subreddit at least 5~times, which is denoted as a
``social home'' in previous work~\cite{datta2019extracting}. We do this to avoid
counting single, spurious toxic comments in one-off communities.

When abusive accounts post toxic comments in their social homes, they often do
so in violation of subreddit norms. Abusive accounts violate the toxicity
norms of a median 16.7\% of their social homes
(Figure~\ref{fig:abuser_behavior_subreddits_cdf}). Again, there is significant
variance in behaviors; 5.8\% of abusive accounts violate the toxicity norms of
every single subreddit they are a part of. When abusive accounts do violate
subcommunity norms, they often do so to significant degrees---abusers post a
median 3.55x more toxic comments than the community standard. Such behaviors suggest
that extreme norm-violations can serve as a broader signal to detect the most toxic
accounts.

\subsection{Classes of abusive accounts}
\begin{table}
    \centering
    \footnotesize
    \begin{tabularx}{\columnwidth}{Xrrrr}
        \toprule
        Metric  &   Sub-Metric  &   Cluster 1   &   Cluster 2   &   Cluster 3 \\
        \midrule
        Cluster     & Size &   350K (71\%)  &   117K (24\%)    &   21K (4.3\%) \\
                    & Toxic Comments &   71\%    &   26\%    &
                    3.1\% \\
        \midrule
        \multirow{2}{*}{Activity}     & Comments &  521 &  144 & 25 \\
                            &  Subreddits &   63 &   22  &  3    \\
                            &  Social Homes &  18 &   5  &   1 \\
        \midrule
        \multirow{2}{*}{Toxicity}   & Agg. Toxicity      &   2.4\%   &   7.2\%
        &   18\% \\
                                    & Tox Subreddits     &   11.4\%  &   25\%
                                    &   50\% \\
                                    & Violat. Subreddits   &   12.5\%  &   40\%
                                    &   100\% \\
        \bottomrule
    \end{tabularx}
    \caption{\textbf{Abuser Personas}---%
        Abusers fall into three distinct \emph{personas} that capture their
        toxicity behaviors on the platform. Cluster 1 abusers are occasional
        abusers, who post only a small handful of toxic comments but are
        otherwise active, regular members of their social communities. Cluster
        2~abusers are moderate abusers who post more toxic comments in more of
        their subcommunities, but are still discerning about their toxicity
        behaviors and more conscious of subcommunity norms. Cluster 3~abusers
        are serial abusers who post a significant fraction of comments and
        regularly violate community norms, suggesting platform wide defenses may
        be the only way to curb abuse from those types of accounts.
    }
    \label{table:abuser_personas}
\end{table}

We leverage our analysis of abusive behaviors to this point to ultimately build
distinct abuser personas. We aggregate four key features of abuser toxicity
behaviors:

\begin{enumerate}
    \item The fraction of comments that are toxic in aggregate.
    \item The fraction of subreddits that contain a toxic comment.
    \item The median fraction of toxic comments for each subreddit the abuser
        participates in.
    \item The fraction of subcommunities that each account violates a toxicity
        norm in.
\end{enumerate}

We cluster abusive accounts using K-Means with Principal Component Analysis
(PCA) for dimensionality reduction. We reduced to three components, as they
capture the majority of the variance between each variable. We note that not
every abusive account has explicit norm violation statistics (given only 35.8K
communities have stable toxicity norms)---we exclude these abusive accounts and
cluster the 489K (53\%) remaining accounts. Abusers fall into three distinct
classes of abusive accounts. Each cluster's median behavior is detailed in
Table~\ref{table:abuser_personas}. We detail each cluster below:

\paragraph{Cluster 1: Occasional abusers}
350K (71.7\%) abusive accounts are occasional abusers, which are accounts that post
relatively small fractions of toxic comments during their posting history
(median 2.4\%) and contribute 70\% of all toxic comments posted to Reddit.
These accounts post toxic content in a relatively small fraction of their
subcommunities (11.4\%) and tend to be regular, contributing members of the
subcommunities they belong in. Indeed, these accounts are also the most active
abusive accounts on the platform, posting a median 521~comments and
participating in a median 18~social homes. While these accounts may have a
proclivity towards toxic behaviors, they may also be the ones most amenable to
nudges or other types of interventions, as toxic behaviors do not make up a
significant portion of their overall platform behaviors.

\paragraph{Cluster 2: Moderate abusers}
117K (24.1\%) abusive accounts are moderate abusers, which are accounts that post
moderate amounts of toxic comments on the platform (median 7.2\% toxic
comments) and contribute 26\% of all toxic comments posted to Reddit. Notably,
these accounts are toxic in a larger fraction of their subcommunities, and
violate the norms of a median 40\% of communities they participate in. This is
approximately 3.5~times that of occasional abusers but still a minority of
their subreddits, highlighting that moderate abusers are selectively
abusive in a handful of communities but not others. Curbing abuse from these
types of accounts is challenging, as simple nudges are likely ineffective.
Instead, more robust defenses that include subcommunity specific moderation
practices may be most effective.

\paragraph{Cluster 3: Serial abusers}
21K (4.3\%) of accounts are serial abusers, which is the least prevalent abuser
persona and contribute just 3.1\% of all toxic comments posted to Reddit.  These
accounts are \emph{serial abusers}---they post a median 18.1\% toxic comments,
and post toxic comments in a median 50\% of the subreddits they participate in.
To make matters worse, they violate the toxicity norms of \emph{every
subcommunity they post toxic content in}, often entirely disregarding
established toxicity norms. Encouragingly, these types of accounts are the least
active type of abuser, limiting their existing impact on the platform. Still,
given their proclivity to posting toxic content, simple interstitial defenses
such as nudges may not be effective in curbing abuse from these accounts.
Further abuse from these accounts can likely only be curbed through
platform-level action (e.g., bans, suspensions).

\section{Case study: The Murder of George Floyd}
\label{section:case_study}
\begin{table}
    \centering
    \small
    \begin{tabularx}{\columnwidth}{Xrr}
        \toprule
        Metric      &   Control     &   George Floyd Incident   \\
        \midrule
        Toxicity Volume &   68K (0.7\%) &   91K (0.9\%) \\
        Abusive Accounts*   &   332K (15.4\%)    &   347K (15.7\%) \\
        Spurious Receivers    &   17.5K (95.1\%) &   23.2K (93.9\%) \\
        Repeat Receivers  &   575 (3.1\%) &   907 (3.7\%) \\
        Flooded Receivers &   315 (1.7\%)     &   584 (2.4\%) \\
        \bottomrule
    \end{tabularx}
    \caption{\textbf{Toxicity in response to the murder of George Floyd}---%
        The fraction of toxic comments increased by 30\% on Reddit in response
        to the murder of George Floyd. This had downstream impacts on receiver
        experiences, which slightly skewed away from spurious toxic interactions
        and towards more impactful forms of abuse, like repeated attacks and
        flooding. Results are statistically significantly different between the
        two groups except when denoted with an asterisk.
    }
    \label{table:case_study_toxicity}
\end{table}

In our longitudinal analysis, we observed several spikes in toxic
behaviors on Reddit. One such spike occurred between May 26th, 2020 and June
5th, 2020, which reached a peak during a three day period between May 29th and
May 31st, 2020. This spike was primarily in response to the murder of George
Floyd. In this section, we detail the impact of this real-world event on toxic
interactions on the platform in aggregate, on abusive account behaviors, and on
receiver experiences. To do this, we compare behaviors from this spike period
against a 3~day sample of the dataset collected from May 1st, 2020 to May 4th,
2020 as a control.



\subsection{Changes in abuser behaviors}
During the peak of toxic behaviors, 0.9\% of all comments on the platform were
toxic, which marks a 30\% increase in overall average toxicity from the control
period (Table~\ref{table:case_study_toxicity}). Overall abuser activity (\eg
number of comments, number of subreddits) stayed consistent throughout the
control period and spike period, suggesting that the increase in toxic comment
volume was not directly related to a significant number of new abusive accounts
becoming active during this period. Rather, we observe that 90.2\% of abusive
accounts posted either the same volume or more \emph{toxic} comments during the
spike period compared to the control period, which contributed overall to an
increased period of toxicity throughout the platform.

We observe no changes in the \emph{structure} of toxic interactions (\eg those
discussed in Section~\ref{section:abuser_relations}), suggesting that such
behavioral patterns remained consistent even when abusive accounts increased
their volume of toxicity. Despite this, receiver experiences shifted slightly
during the spike period. While interactions largely remained spurious, the
number of receivers with solely spurious interactions decreased from 95.1\% to
93.9\%, and instead shifted towards more impactful forms of abuse, like repeated
abuse (3.7\%) and flooding (2.4\%). This skew towards repeated abuse and
flooding was largely in discussions of the George Floyd incident. For example,
one account posted in \texttt{r/PublicFreakout}:

\begin{quote}
    ``Now those *same exact people* are defending what these fascist pigs are
    doing.''
\end{quote}

\noindent The comment was met with 6~different attacks berating them for their comment and
insulting them. The slight shift in the types of attacks that receivers
experienced during the spike of toxicity may anecdotally suggest that the types
of toxic interactions may increase in intensity during heated discussion of
real-world events.

\subsection{Subcommunity spread}
Despite abuser behaviors remaining relatively consistent, we observed that the
subcommunities where toxic behaviors took place changed during the spike period.
Many large subreddits that were closely discussing the incidents as well as the
resulting protests boomed in posting volume. As an example, \texttt{r/PublicFreakout}, which is a community designed for discussing videos
of ``people freaking out, melting down, losing their cool, or being weird in
public''\footnote{\url{https://www.reddit.com/r/PublicFreakout/}}, saw its
comment volume increase by 620\% and its toxic comment volume increase by 670\%
during the spike. This also impacted many smaller subcommunities---for example,
the subreddit with the largest change in toxic comment volume (9566\% increase
in toxic comments) was \texttt{r/Minneapolis}, which is where the murder of
George Floyd took place. Other communities with a stark increase in toxic
comments were other cities where protests were taking place, for example,
\texttt{r/philadelphia} (3720\% increase), and \texttt{r/cincinnati} (3000\%
increase).

In all of these cases, we observe that many toxic comments are posted by
accounts that \emph{never post in the subcommunity prior to the event}. 569
(18.2\%) of the accounts that posted toxic comments in \texttt{r/PublicFreakout}
never posted in this subreddit prior. A similar result holds true for
\texttt{r/Minneapolis} (26.2\% new members), \texttt{r/philadelphia} (24.5\%)
and \texttt{r/cincinnati} (10.1\%), suggesting that at least some fraction of an
increase in toxic content in these subcommunities comes from \emph{outsider}
accounts, likely joining these subcommunities to discuss ongoing incidents and
post inflammatory content. As an example, one new account that joined
\texttt{r/Minneapolis} posted inflammatory responses to accounts talking about
the ongoing protests. In one instance, they wrote:

\begin{quote}
    ``hmm, lets ruin people's businesses, earnings they have worked for to feed
    their family and shit, or work that they have put years into being ruined.
    yeah you are a bunch of fucking retards, all of you need to be executed''
\end{quote}

Such examples highlight that some abusive accounts may actively seek out
contentious discussion and participate in a toxic manner during known real-world
events on the platform.

\section{Discussion and Limitations}
In this section, we synthesize our contributions into a set of open challenges and research directions for improving the automated detection of toxic behaviors and empowering community governance.

\subsection{Abusive Accounts Contribute Significantly to Reddit}
Toxicity on Reddit accounts for a relatively small fraction of comments but are highly visible on the platform: 55.2\% of Reddit accounts post directly on a thread with a toxic comment. Abusive accounts themselves make up just 3.1\% of all accounts, but make up 33.3\% of all comments to Reddit, which aligns with studies on other platforms; Hindman et~al. describe the challenge on Facebook as a ``superuser supremeacy problem~\cite{fb-superuser}.'' The majority of these accounts engage in abusive infractions throughout their lifetime, but simultaneously contribute significant volumes of non-toxic content, ostensibly making them valuable contributors to the platform at large. As such, traditional strategies for dealing with abusive accounts (\eg mass account bans) are likely untenable, as they would significantly reduce meaningful conversation on the platform and have potentially unforeseen consequences on platform health. Some platforms are attempting more nuanced actions. For example, Twitter deployed a strike system for handling accounts that post Covid-19 misinformation~\cite{twitter-strike}, and some subreddits have implemented similar strike systems for moderation at large~\cite{programmer-strike}. Still, there is limited insight into how effective these strategies may be for handling online hate and harassment. The design of new defensive schemes and evaluating their efficacy for these types of attacks is a potential direction for future research.

\subsection{Time-based and Graph-based Features can Improve Toxicity Detection}
We uncovered three distinct attack patterns when studying the graph relationships between abusive accounts and their targets: one-off attacks, repeated toxicity, and flooding. In doing so, we also uncovered several graph- and time-based features that may prove useful in providing context around toxic behaviors. For example, prior relationshps between accounts, assortatitivy of abusive accounts (whereby clusters of abusive accounts coordinate or interact frequently) and the target selection of abusive accounts(such as repeatedly sending toxic comments to a single account) provide a richer context compared to isolated attacks. These graph-based and time-based features represent a potential new direction for detecting abusive accounts, and can add additional context to existing classification systems, which rely \emph{exclusively} on text content to arrive
on decisions. Incorporating these features into more comprehensive classifiers is a promising area of future work.

\subsection{Classes of Abusive Accounts can Inform Defensive Design}
Our identification of three
distinct classes of abusive accounts (\eg occasional abusers, serial abusers) and varied community norms suggests the need for targeted interventions, rather
than a one-size-fits-all approach to actioning toxic behaviors. As we previously
proposed, one-off attackers might benefit most from inline warnings or nudges.
Chang \etal previously found that temporarily blocking Wikipedia contributors
substantially reduced the rate of repeated abuse~\cite{chang2019trajectories}.
Likewise, Instagram now includes a feature that warns users before they post
content that appears similar to previously reported hate and
harassment~\cite{instagram-bullying}. However, the existence of moderate and
serial attackers requires more serious interventions, up to and including
suspension or permanent blocking. Chandrasekharan \etal previously showed that
banning subcommunities can be highly effective~\cite{chandrasekharan2017you}.
However, the throw-away nature of accounts on Reddit may complicate applying
such a strategy to individual attackers---though this limitation may not exist
for all online social networks.

At the same time, some communities like \texttt{r/WallStreetBets} and
\texttt{r/RoastMe} relish in offensive, profanity-laced discussions with other
willing participants. Combined with varying personal definitions of what
constitutes toxic content~\cite{kumar2021designing, gordon2021disagreement}, it
is critical that platform designers consider empowering community-level
moderators to best support conversational nuance online. However, it remains
critical to enforce site-wide policies against hate and harassment, lest toxic
subcommunities flourish that negatively impact other communities or
users~\cite{chandrasekharan2017you}.

\subsection{Limitations}
Our work is not without limitations. For one, as we note in Section~\ref{section:methodology}, leveraging the Perspective API alone for toxic comment detection is noisy, and can lead to both false positives and false negatives. We stress that we evaluated our usage of the Perspective API on Reddit comments to provide a fair assessment of its performance throughout our study. Another key limitation is the existence of sockpuppet Reddit accounts, whereby individual users create multiple accounts for different purposes (\eg to spread toxic comments). While some techniques do exist to identify sockpuppets~\cite{kumar2017army,weerasinghe2022using}, these studies are still nascent and can also lead to significant false positives or negatives. While we take some precautions to account for these in our abusive account categorization (\eg restricting to active subreddits, measuring norms from social homes), we note that the presence of sockpuppets may erroneously count multiple accounts as single individuals. Still---we argue that this perspective is the closest aligned to both community moderators and platform policy makers (who can only operate on an account level), and thus our results are still applicable for their use cases.
\section{Conclusion}
In this work, we presented the results of a longitudinal measurement study of
abusive accounts posting toxic comments on Reddit. By pairing the Perspective
API with different model thresholds and curated filters, we identified over
929K~abusive accounts and 14~million toxic comments over an 18~month period.
These accounts engaged in a variety of toxic behaviors, the most prominent of
which were insults, followed by identity attacks, calls to leave the platform,
and threats of violence. We examined a variety of features associated with each
account---such as overall toxicity rates, toxicity rates per community, and
community norm violations---and identified three distinct classes of abusive
accounts that each would benefit from a nuanced intervention. Similarly, we
explored the graph relationships between attackers and targets and found
multiple examples of mob-like attacks and long-term attack campaigns. Our
measurements serve to better understand the dynamics of hate and harassment
attacks in practice, as well as to identify features that might transform
classification from a per-comment decision into a holistic time-based,
graph-based, and content-based assessment.


{
\balance
\bibliographystyle{ACM-Reference-Format}
\bibliography{paper}
}
\clearpage
\appendix
\section{Evaluating Abuser Thresholds}
\begin{table}[t]
    \centering
    \small
    \begin{tabularx}{\columnwidth}{Xrr}
        \toprule
        Comment Threshold   &   \# Abusive Accounts &   Precision \\
        \midrule
        1           &   931K (2.9\%)    &   0.71 \\
        2           &   332K (1\%)      &   0.68 \\
        3           &   183K (0.5\%)    &   0.68 \\
        4           &   119K (0.4\%)    &   0.7 \\
        5           &   84K  (0.3\%)    &   0.66 \\
        \bottomrule
    \end{tabularx}
    \caption{\textbf{Abusive Account Thresholds}---%
        Increasing the number of toxic comments required to label an account as
        an abusive account does not demonstrably improve performance, while
        significantly reducing the account and comment volume available to
        study.
    }
    \label{table:abuser_thresholds}
\end{table}

\label{section:abuser_thresholds} For the scope of this study, we consider an
account to be abusive if it posts just a \emph{single} comment above the high
precision threshold of \texttt{SEVERE\_TOXICITY} $>0.9$. However, we also
evaluated whether increasing the number of high-precision toxic comments
required to label an account as abusive could in turn increase our overall
precision. To measure this, one expert rater manually sampled high-precision
toxic comments from accounts that posted 1--5 toxic comments, and we evaluated
the resultant precision to see if changing the threshold for abusive comments
would increase our results. Table~\ref{table:abuser_thresholds} shows the
results. Ultimately, precision was stable for all samples at 0.7, suggesting
that increasing the threshold would not result in higher quality data
while also reducing the size of the abusive account population to a tenth of the
size in the most extreme case.

\end{document}